\documentclass[aps,prl,10pt,floatfix,tightenlines,amsmath,amssymb,showpacs,superscriptaddress,nofootinbib,twocolumn]{revtex4-2}
\usepackage{amsmath}
\usepackage{graphicx}
\usepackage{epstopdf}
\usepackage{dcolumn}
\usepackage{color}
\usepackage{bm}
\usepackage{overpic}
\usepackage{rotating}
\usepackage{times}
\usepackage{multirow}
\usepackage{float}
\usepackage{mathrsfs}
\usepackage{booktabs}
\usepackage{array,booktabs}
\usepackage[mathlines]{lineno} 
\usepackage{epstopdf}
\usepackage{hyperref}
\usepackage{amssymb}
\usepackage{graphicx}
\newcommand{\ee}{e^+e^-}

\newcommand{\DDpi}{D^0D^-\pi^+}

\begin{document}
	
	\lefthyphenmin=2
	\righthyphenmin=2
	\tolerance=1000
	\uchyph=0
	
	\normalsize
	\parskip=5pt plus 1pt minus 1pt
	
\title{\boldmath Observation of a Charged Charmoniumlike Structure in $\ee\to\DDpi+c.c.$}

\newcommand{\BESIIIorcid}[1]{\href{https://orcid.org/#1}{\hspace*{0.1em}\raisebox{-0.45ex}{\includegraphics[width=1em]{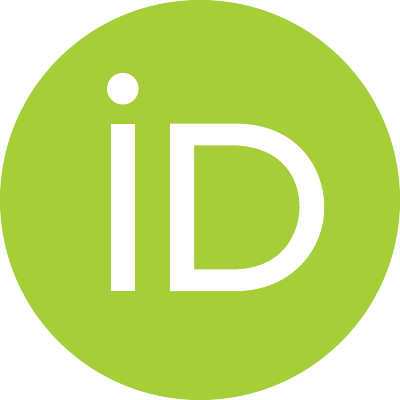}}}} 
	
\author{
M.~Ablikim$^{1}$\BESIIIorcid{0000-0002-3935-619X},
M.~N.~Achasov$^{4,c}$\BESIIIorcid{0000-0002-9400-8622},
P.~Adlarson$^{83}$\BESIIIorcid{0000-0001-6280-3851},
X.~C.~Ai$^{89}$\BESIIIorcid{0000-0003-3856-2415},
C.~S.~Akondi$^{31A,31B}$\BESIIIorcid{0000-0001-6303-5217},
R.~Aliberti$^{39}$\BESIIIorcid{0000-0003-3500-4012},
A.~Amoroso$^{82A,82C}$\BESIIIorcid{0000-0002-3095-8610},
Q.~An$^{79,65,\dagger}$,
Y.~H.~An$^{89}$\BESIIIorcid{0009-0008-3419-0849},
Y.~Bai$^{63}$\BESIIIorcid{0000-0001-6593-5665}
% ...
\\
\vspace{0.2cm}
(BESIII Collaboration)\\
\vspace{0.2cm} {\it
$^{1}$ Institute of High Energy Physics, Beijing 100049, People's Republic of China\\
% ...
$^{4}$ Budker Institute of Nuclear Physics SB RAS (BINP), Novosibirsk 630090, Russia\\
% ...
$^{31}$ INFN Sezione di Ferrara, (A)INFN Sezione di Ferrara, I-44122, Ferrara, Italy; (B)University of Ferrara, I-44122, Ferrara, Italy\\
% ...
$^{39}$ Johannes Gutenberg University of Mainz, Johann-Joachim-Becher-Weg 45, D-55099 Mainz, Germany\\
% ...
$^{63}$ Southeast University, Nanjing 211100, People's Republic of China\\
% ...
$^{65}$ State Key Laboratory of Particle Detection and Electronics, Beijing 100049, Hefei 230026, People's Republic of China\\
% ...
$^{79}$ University of Science and Technology of China, Hefei 230026, People's Republic of China\\
% ...
$^{82}$ University of Turin and INFN, (A)University of Turin, I-10125, Turin, Italy; (B)University of Eastern Piedmont, I-15121, Alessandria, Italy; (C)INFN, I-10125, Turin, Italy\\
$^{83}$ Uppsala University, Box 516, SE-75120 Uppsala, Sweden\\
% ...
$^{89}$ Zhengzhou University, Zhengzhou 450001, People's Republic of China\\
\vspace{0.2cm}
$^{\dagger}$ Deceased\\
% ...
$^{c}$ Also at the Novosibirsk State University, Novosibirsk, 630090, Russia\\
% ...
}}

	
\begin{abstract}
We present a partial wave analysis of the process $\ee \to D^0 D^- \pi^++c.c.$ using a data sample collected with the BESIII detector at a center-of-mass energy of 4682~MeV. 
A charged structure, referred to as the $X(3790)^-$ near the $D^0D^-$ mass threshold with a significance of $5.3\sigma$, which includes systematic uncertainties and the look-elsewhere effect. 
Assuming a Flatté description of the structure, the quantum numbers of the $X(3790)^-$ are found to be $J^{P}=1^{-}$, and its mass and coupling strength to the $D^0D^-$ channel are determined to be $(3794.7 \pm 6.0 \pm 8.7)$~MeV/$c^2$ and $(1.6 \pm 0.2 \pm 0.2)$~GeV, respectively, where the first uncertainties are statistical and the second systematic. 
In addition, we also report measurements of the masses and widths of the $D_{2}^{*}(2460)$, $D_{1}^{*}(2600)$, and $D_{1}^{*}(2760)$ charmed mesons.

\end{abstract}
	
	
\maketitle

The $X(3872)$ was first observed by the Belle Collaboration in 2003 in the $\pi^+\pi^- J/\psi$ invariant mass spectrum from $B^+\to K^+ \pi^+\pi^- J/\psi$ decays~\cite{finding_X_3872}. 
Its mass lies extremely close to the $D^0\bar{D}^{*0}$ threshold, and its narrow width, $J^{PC}=1^{++}$ quantum numbers \cite{jpc_X_3872}, and isospin-violating decays~\cite{finding_X_3872, X_3872_rho_jpsi, X_3872_omega_jpsi1, X_3872_omega_jpsi2, ratio_X_38721, ratio_X_38722, wangxf} challenge a conventional charmonium interpretation. 
These features have stimulated various theoretical model calculations, and there is broad consensus that there is a large $D^0\bar{D}^{*0}$ molecular component in its wave function~\cite{ratio_X_38721, ratio_X_38722, mole_X_3872}.
Additionally, several charged charmoniumlike structures, such as the $Z_c(3900)$ and $Z_c(4020)$ were discovered 
accompanied by a pion meson in $e^+e^-$ annihilation~\cite{Zc_1, Zc_2, Zc_4, Zc_5, Zc_6}. Their masses are also very close to the thresholds of a pair of charmed mesons, indicating they may also contain large $D^{(*)}\bar{D}^{(*)}$ molecular components.  

In the hadronic molecular picture, heavy quark symmetry implies the existence of partner states formed from $D^{(*)}\bar{D}^{(*)}$ pairs~\cite{Guo2013PRDSpinPartner}. 
Experimental searches for near-threshold scalar enhancements in final states such as $D^0\bar{D}^0$ and $D^{*0}\bar{D}^{*0}$ have been carried out~\cite{X_3872_sp_1, X_3872_sp_2, X_3872_sp_3}. However, no dedicated studies have yet been reported for isovector states in the $(D\bar{D})^\pm$ final state.
Such states were previously considered unlikely, as spin-parity conservation forbids the $D\bar{D}\pi$ vertex under the assumption that hadronic molecules are predominantly bound via pion exchange, analogous to nuclear forces. Nevertheless, phenomenological models and lattice QCD simulations predict the existence of a scalar $D\bar{D}$ bound state~\cite{DDbar_predict_1, DDbar_predict_2}.

The process $e^+e^-\to \DDpi$ \footnote{Including $D^0D^-\pi^+$ and $\bar{D}^{0}D^{+}\pi^-$, charge conjugated modes are implied throughout unless stated otherwise.} offers a valuable opportunity to search for such isovector states coupling to $D\bar{D}$. It also serves as an ideal reaction in which to measure the parameters of the excited charmed mesons with non-exotic quantum numbers, such as the $D_2^*(2460)$, $D_1^*(2600)$, and $D_1^*(2760)$, which decay into the $D\pi$ final state. 
These excited charmed mesons have been observed over the past decades by \textit{BABAR}, Belle, CLEO, and LHCb~\cite{pdg}. Their properties have motivated extensive theoretical efforts aimed at classifying these states~\cite{HQL_D, ex_D_1, ex_D_2, ex_D_3, ex_D_4, ex_D_5, ex_D_6, ex_D_7}.
Nevertheless, precise experimental input on resonance parameters, as well as on production and decay mechanisms, remains essential for drawing firm conclusions about their nature.

This Letter presents a partial wave analysis (PWA) of the process $\ee\to \DDpi$ at a center-of-mass energy $\sqrt{s} = 4682$~MeV~\cite{ecms_lum}. 
The analysis is based on a data sample corresponding to an integrated luminosity of 1667~pb$^{-1}$~\cite{ecms_lum}, collected with the BESIII detector at the BEPCII collider~\cite{bes3_1, bes3_2}.
Simulated samples, which are used to estimate the background, perform PWA, and implement initial state radiation (ISR) corrections, are produced with {\sc geant4}-based~\cite{geant4} Monte Carlo (MC) software, which includes the geometric description of the BESIII detector and its response. The process $\ee\to \DDpi$ is simulated using a phase space (PHSP) model~\cite{evtgen_1, evtgen_2}. The decays $D^0\to K^-\pi^+$ and $D^-\to K_S^0\pi^-$ are simulated using the PHSP model as well. The decays $D^0\to K^-\pi^+\pi^0$, $D^-\to K^+\pi^-\pi^-$, and $D^-\to K_S^0\pi^-\pi^0$ are simulated using the D\_Dalitz model~\cite{evtgen_1, evtgen_2, D_Dalitz}. The decays $D^0\to K^-\pi^+\pi^+\pi^-$, $D^-\to K^+\pi^-\pi^-\pi^0$, and $D^-\to K_S^0\pi^-\pi^-\pi^+$ are simulated based on PWA results reported in Refs.~\cite{D_K3pi, Dm_K3pi_1, Dm_K3pi_2}. Beam-energy spread and ISR are considered with the generator {\sc kkmc}~\cite{kkmc1, kkmc2}.

We employ both single-tag ($\rm ST$) and double-tag ($\rm DT$) reconstruction methods to identify the $\DDpi$ final states. 
In the $\rm ST$ method, only the charged pion coming directly from the $e^+e^-$ annihilation ($\pi_d^+$) and either a $D^0$ or a $D^-$ meson are reconstructed. The ST samples are categorized by the $D$-meson tag mode as ${\rm ST}^{0}_{K\pi}$ ($D^0 \to K^-\pi^+$), ${\rm ST}^{0}_{K3\pi}$ ($D^0 \to K^-\pi^+\pi^+\pi^-$), and ${\rm ST}^{-}_{K2\pi}$ ($D^- \to K^+\pi^-\pi^-$). 
In the $\rm DT$ method, all the final state particles, $\pi_d^+$, $D^0$, and $D^-$, are reconstructed, with no additional charged tracks allowed in an event. The $D^0$ is reconstructed via $K^-\pi^+$, $K^-\pi^+\pi^+\pi^-$, and $K^-\pi^+\pi^0$ decays, and the $D^-$ via $K^+\pi^-\pi^-\pi^0$, $K_S^0\pi^-$, $K_S^0\pi^-\pi^-\pi^+$, $K^+\pi^-\pi^-$, and $K_S^0\pi^-\pi^0$ decays. 

The selection criteria for charged tracks and photons, and the reconstruction of $\pi^0$ and $K_S^0$ candidates are described in Refs.~\cite{tracks, pi0_Ks}. The charged tracks used to reconstruct the signal process are required to originate from the interaction point. For $K_S^0$ candidates, the pseudo-candidate obtained from a secondary vertex fit~\cite{svtx} is used instead of the individual tracks, and the vertex fit $\chi^2$ ~\cite{svtx} is required to be less than 100.

The backgrounds from $\ee\to D^{*+}D^{-}$ are removed by requiring the $D^0\pi_d^+$ invariant mass [$M(D^0\pi_d^+)$], or equivalently the mass recoiling against the $D^-$~[$RM(D^-)$], to be larger than 2.04~GeV/$c^{2}$.
For the ${\rm ST}^{0}_{K3\pi}$ sample, combinatorial backgrounds from $\ee\to \DDpi$ and $\ee\to \pi^+\pi^-D^+D^-$ are suppressed by requiring the invariant masses of $K^-\pi_d^+\pi_{1/2}^+\pi^-$ [$M(K^-\pi_d^+\pi_{1/2}^+\pi^-)$] and $K^-\pi_d^+\pi_{1/2}^+$ [$M(K^-\pi_d^+\pi_{1/2}^+)$] to satisfy $|M(K^-\pi_d^+\pi_{1/2}^+\pi^-) - m_{D^0}| > 10$~MeV/$c^2$ and $|M(K^-\pi_d^+\pi_{1/2}^+) - m_{D^+}| > 16$~MeV/$c^2$, respectively, where the two $\pi^+$s from $D^0\to K^-\pi^+\pi^+\pi^-$ are assigned as $\pi_{1/2}^+$.
For the $\rm DT$ sample, the total four-momentum of the $\pi_d^+$ and the reconstructed $D^0$ and $D^-$ mesons is constrained to that of the initial $\ee$ system by a kinematic fit; the fit must converge with a $\chi^{2}<40$. For ST samples, the invariant masses of $K^-\pi^{+}$ and $K^-\pi^+\pi^+\pi^-$ are constrained to the nominal mass of the $D^{0}$, $m_{D^0}$~\cite{pdg}, and that of $K^+\pi^-\pi^-$ to $m_{D^-}$~\cite{pdg}.

In both the $\rm ST$ and $\rm DT$ samples, the signal and sideband regions for the $D^0$ and $D^-$ candidates are defined using elliptical selection criteria, as detailed in the Supplemental Material~\cite{supp}. 
We define an $S$ sample from events in the signal region, and a $B$ sample, used for background estimation, from events in the sideband regions. 
To improve mass resolution in the ${\rm DT}$ sample, the invariant masses of the $D^0$ and $D^-$ are constrained to $m_{D^0}$ and $m_{D^-}$, respectively. In the $\rm ST$ samples, the recoiling mass of the $D^0\pi_d^+$ or $D^-\pi_d^+$ [the unreconstructed $D$ meson ($D_{\rm miss}^{-}$ or $D_{\rm miss}^{0}$) in the ${\rm ST}^{0}_{K\pi, K3\pi}$ or ${\rm ST}^{-}_{K2\pi}$ samples] is constrained to $m_{D^-}$ or $m_{D^0}$, respectively. The total four momentum of the $\pi_d^+$, the reconstructed $D$ meson, and $D_{\rm miss}$ is constrained to that of the initial $\ee$ system.

\begin{figure*}[!htbp]
    \centering
    \includegraphics[width=0.33\textwidth]{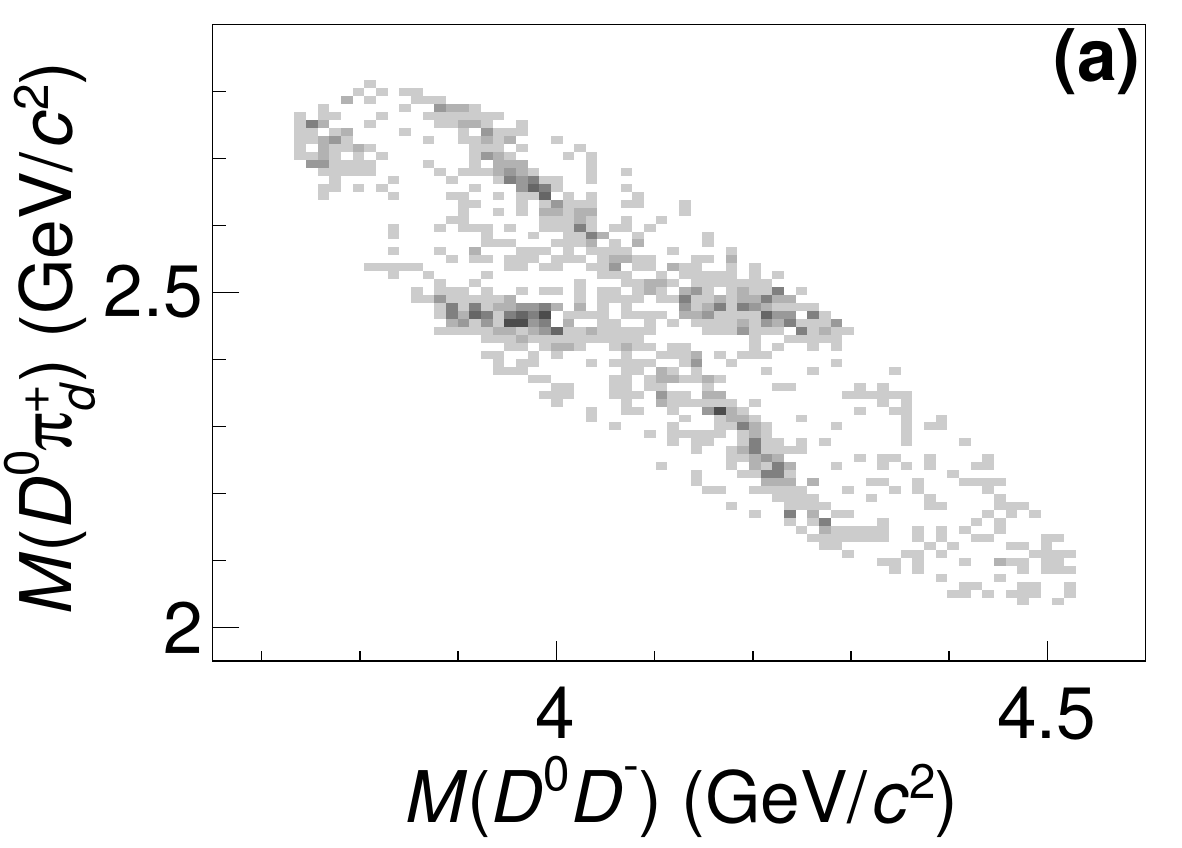}
    \includegraphics[width=0.33\textwidth]{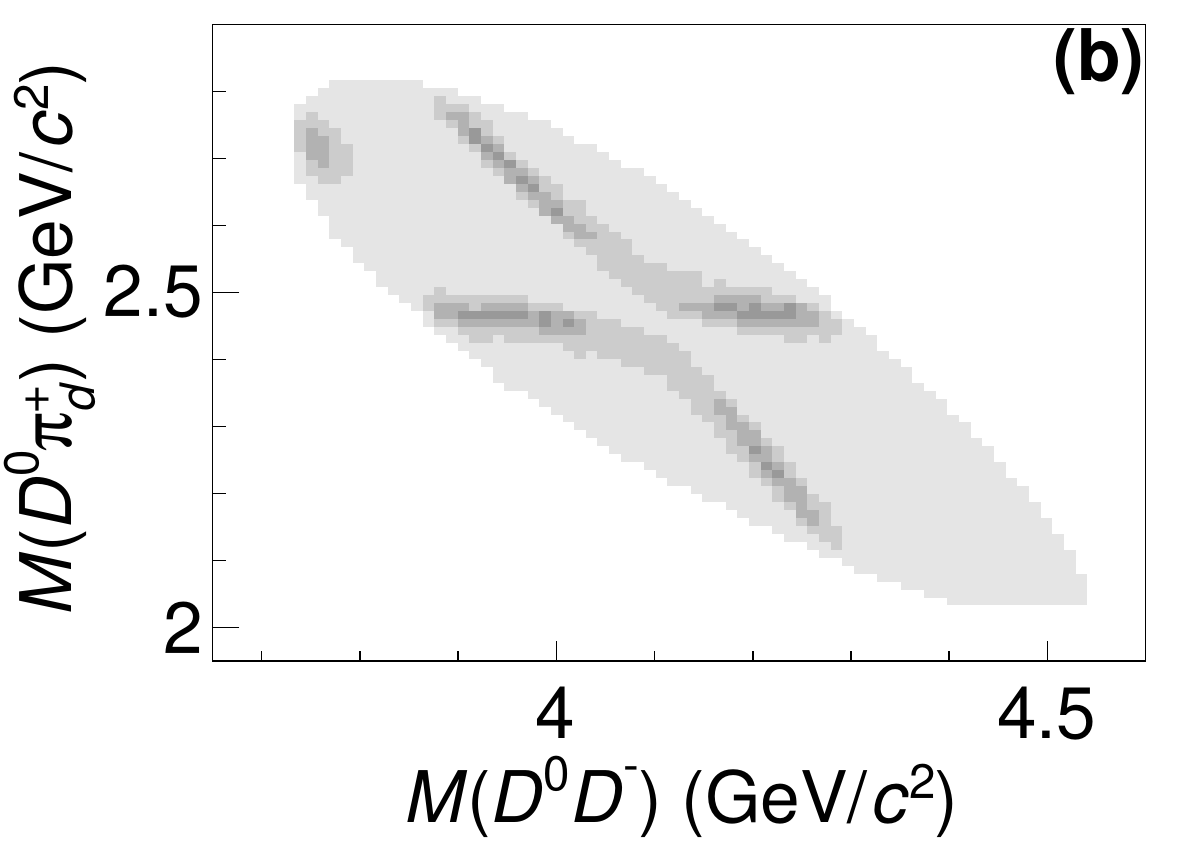}
    \caption{Dalitz plots of selected signal events (a) and the PWA fit (b) from the $\rm DT$ sample. The intensities in (a) and (b) indicate the number of events, with the same scale ranging from 0 to 9.}
    \label{pic:pwa_base_dalitz}
\end{figure*}

After applying the above selection criteria, the Dalitz plots 
of the selected events in the $\rm DT$ sample are shown in Fig.~\ref{pic:pwa_base_dalitz}. 
The mass projections of the selected events in the $\rm DT$ sample together with the $\pi_d^+$ recoil mass [$RM(\pi_d^+)$] distributions of the $\rm ST$ samples are shown in Fig.~\ref{pic:pwa_base} (other distributions are shown in the Supplemental Material~\cite{supp}). 
Clear $D_2^*(2460)^+$ and $\bar{D}_2^*(2460)^0$ bands in the $D\pi$ system are seen in the Dalitz plots. In addition, a clear enhancement near the $D^0D^-$ mass threshold is observed in the $D^0D^-$ invariant mass [$M(D^0D^-)$] distribution of the $\rm DT$ sample, as well as in the $RM(\pi_d^+)$ distributions of the $\rm ST$ samples. 

\begin{figure*}[!htbp]
    \centering
    \includegraphics[width=0.33\textwidth]{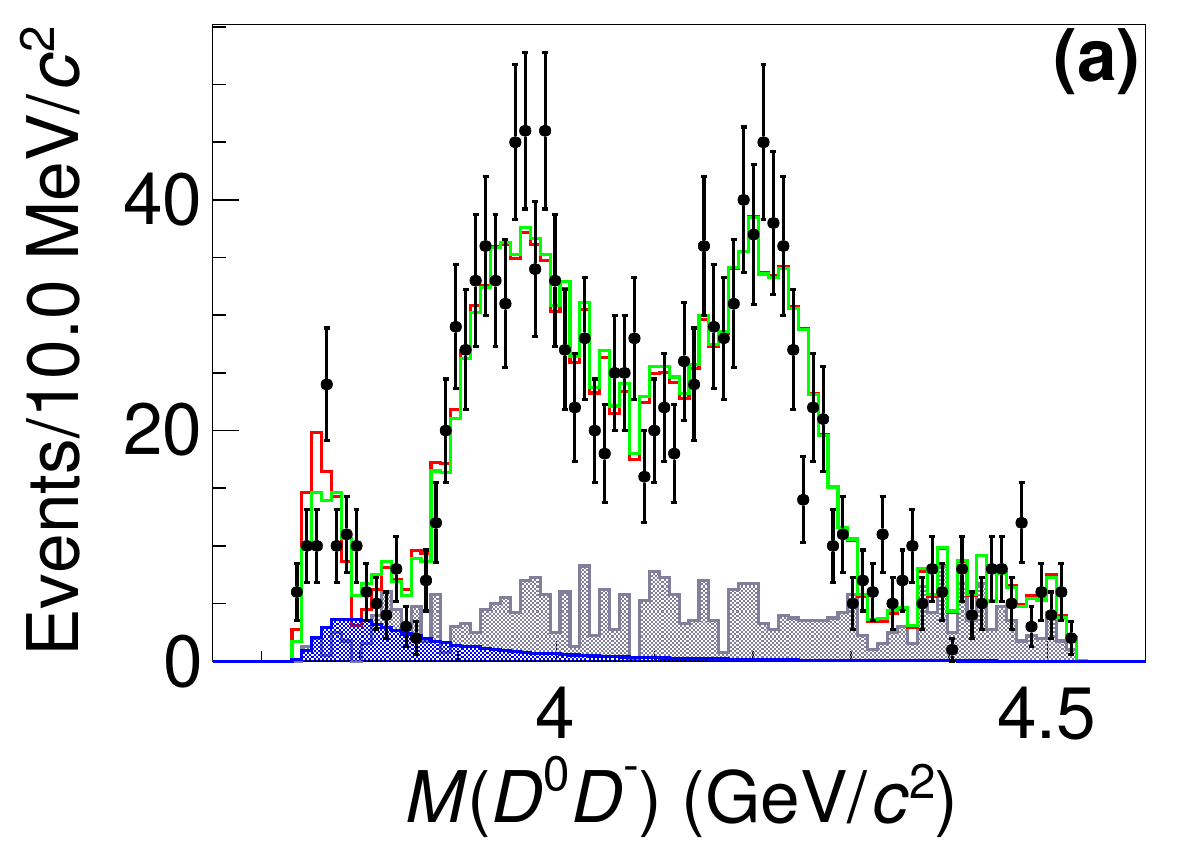}
    \includegraphics[width=0.33\textwidth]{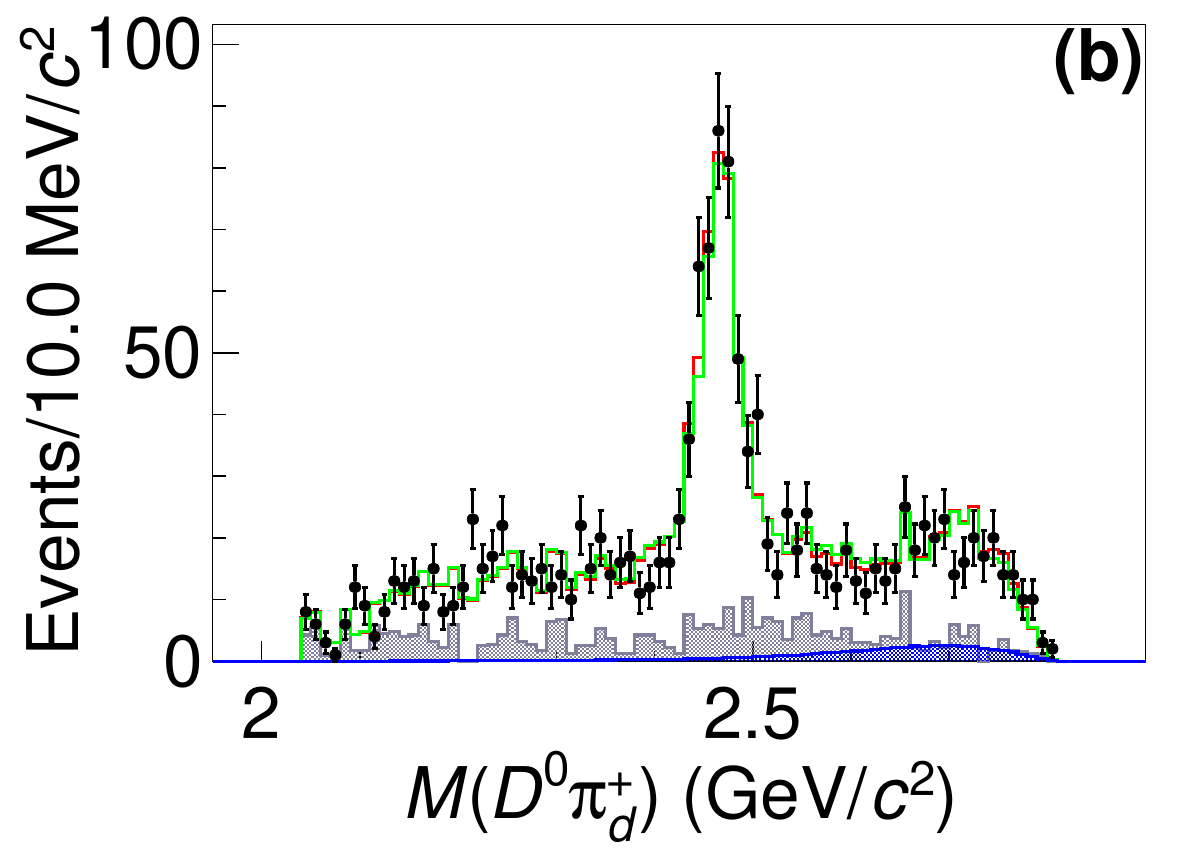}
    \includegraphics[width=0.33\textwidth]{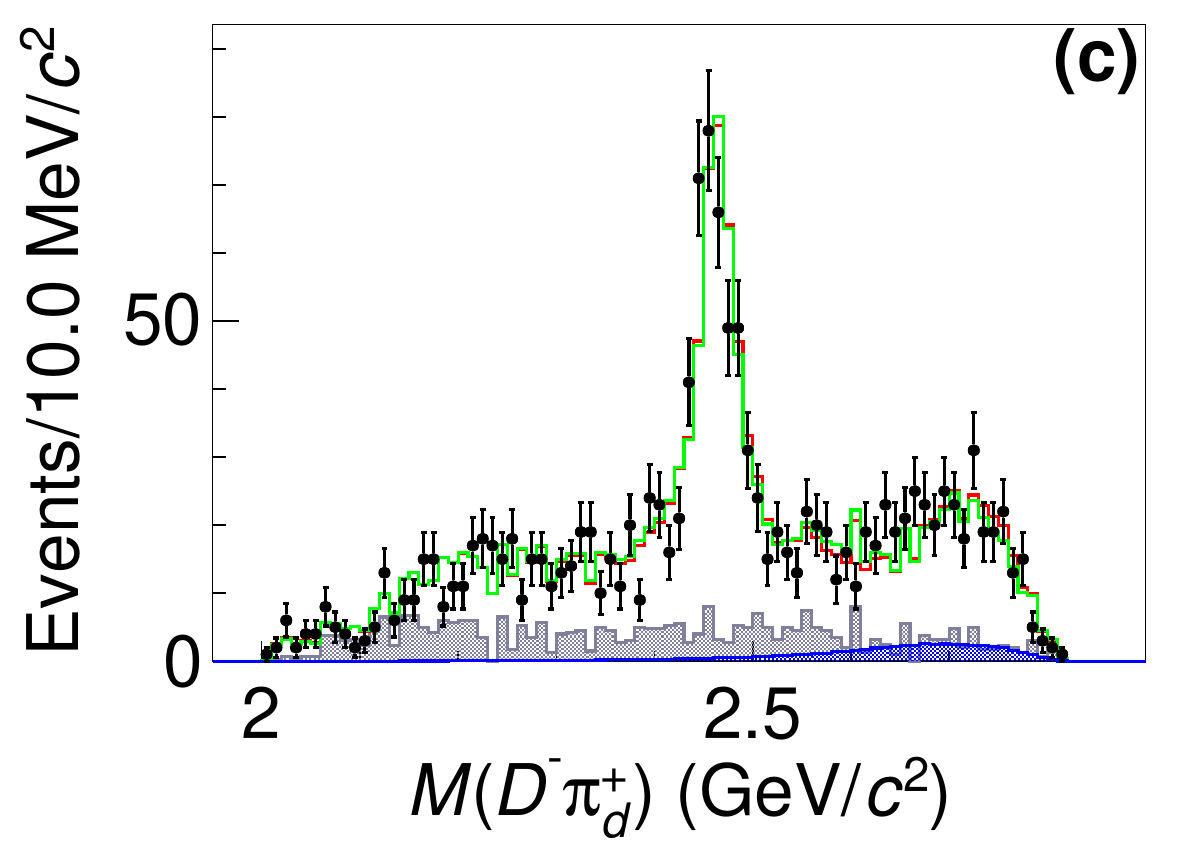} \\
    \includegraphics[width=0.33\textwidth]{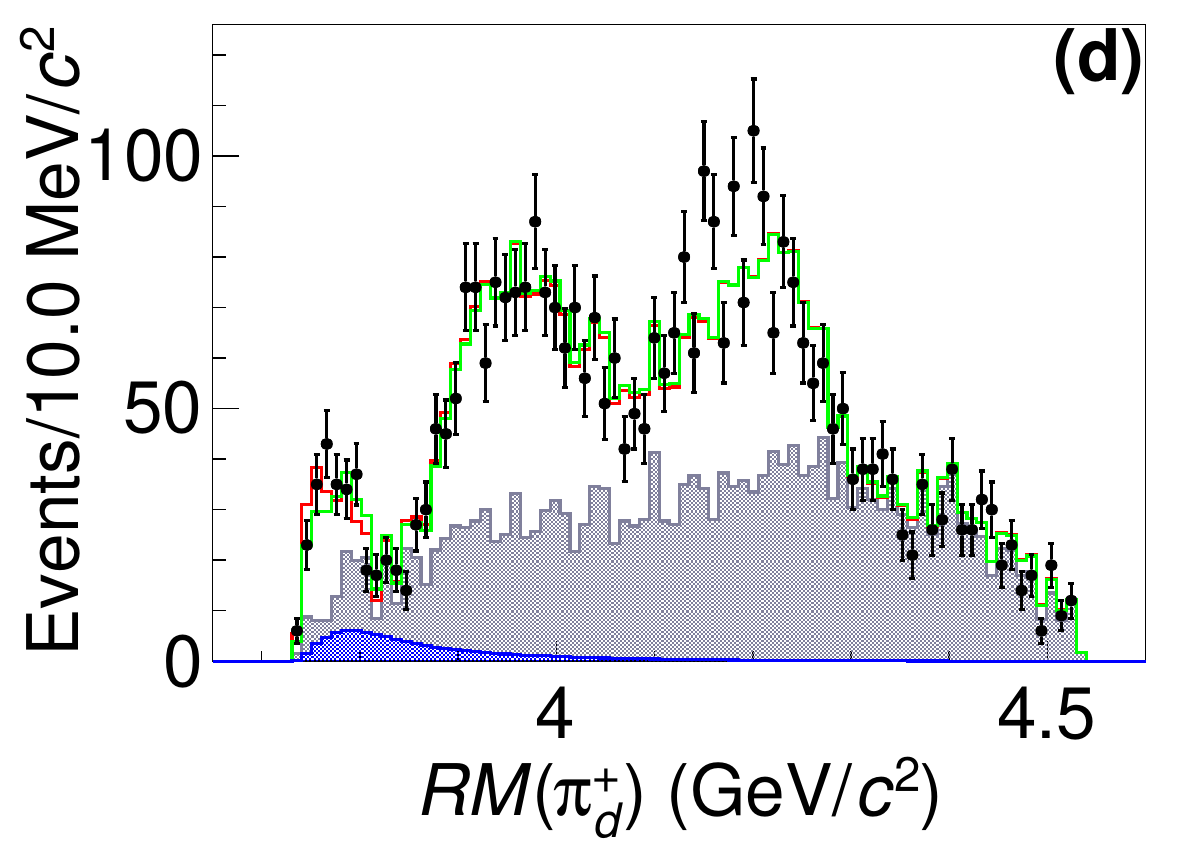}
    \includegraphics[width=0.33\textwidth]{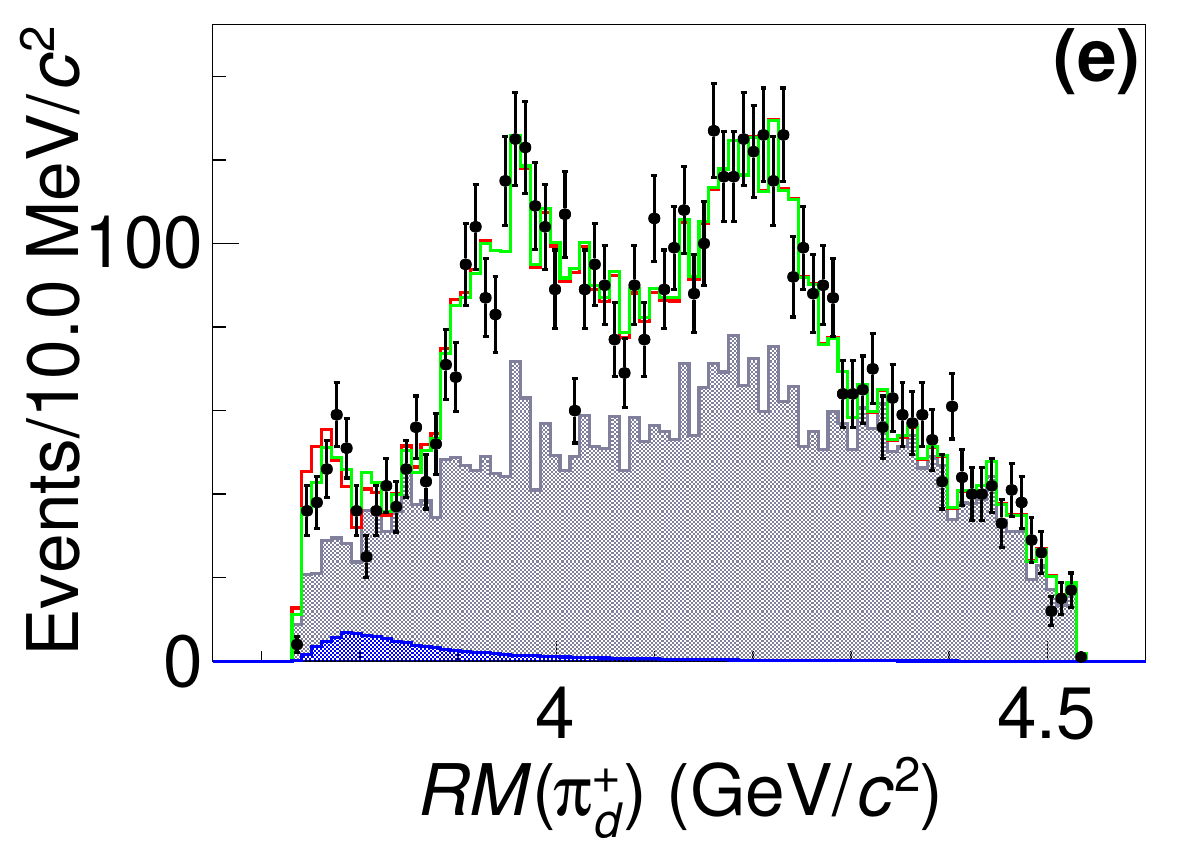}
    \includegraphics[width=0.33\textwidth]{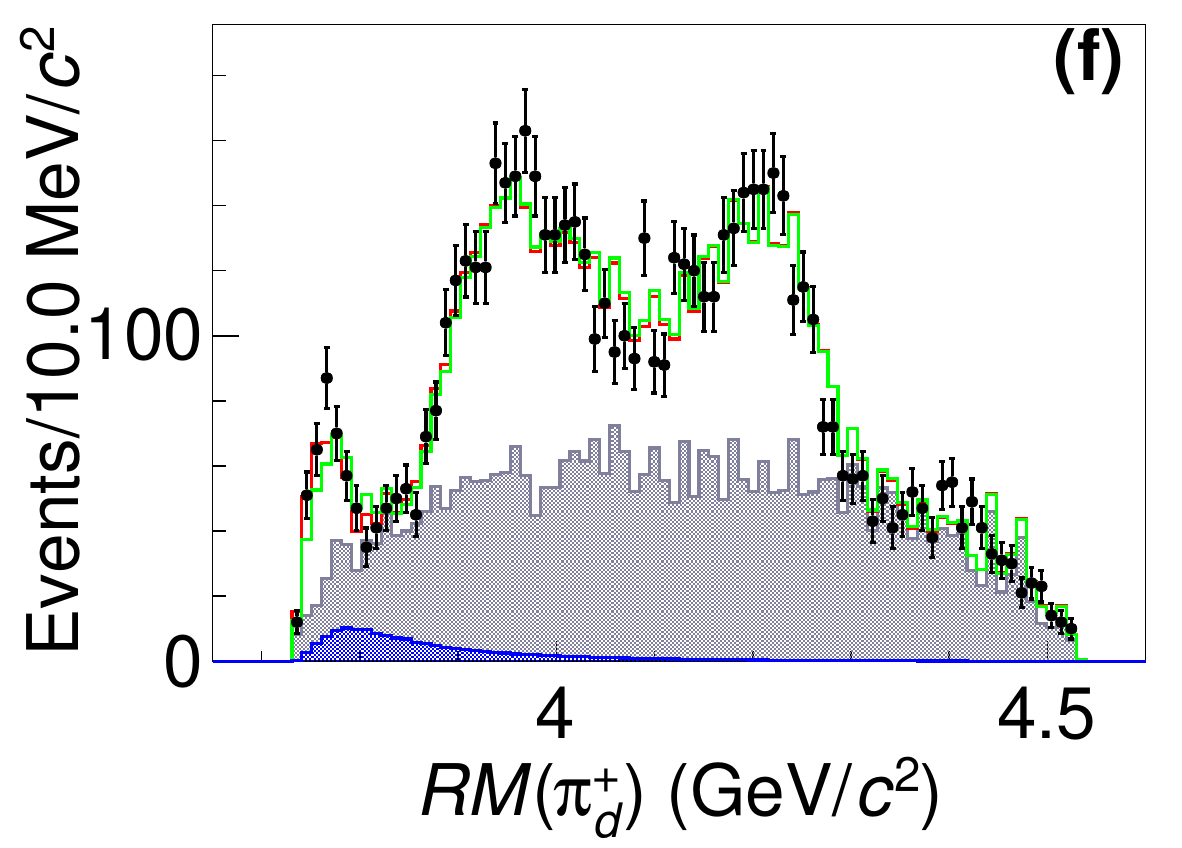}
    \caption{
    Mass projections of $M(D^0D^-)$ (a), $M(D^0\pi_d^+)$ (b), and invariant mass of $D^-\pi_d^+$ [$M(D^-\pi_d^+)$] (c) from the $\rm DT$ sample, and $RM(\pi_d^+)$ in the ${\rm ST}^{0}_{K\pi}$ (d), ${\rm ST}^{0}_{K3\pi}$ (e), and ${\rm ST}^{-}_{K2\pi}$ (f) samples. The black dots with error bars represent the $S$ sample. The red (green) histogram corresponds to PWA fit projections with (without) the $X(3790)^-$ component. The grey and blue histograms are the $B$ sample and the $X(3790)^-$ component, respectively. 
    }
    \label{pic:pwa_base}
\end{figure*}

A PWA is performed on the selected ${\rm ST}$ and ${\rm DT}$ samples using the TF-PWA framework~\cite{tfpwa}, with overlapping events in the ${\rm ST}$ and ${\rm DT}$ samples removed from the ${\rm ST}$ samples to avoid double counting. For the ${\rm DT}$ sample, the four-momenta of the $\pi_d^+$, $D^0$, and $D^-$ are used, whereas for the ${\rm ST}$ samples, the four-momenta of the $\pi_d^+$, the reconstructed $D$ meson, and the $D_{\rm miss}$ are used.
An unbinned maximum likelihood fit is performed by summing the negative log-likelihood (NLL) contributions from the ${\rm DT}$, ${\rm ST}^{0}_{K\pi}$, ${\rm ST}^{0}_{K3\pi}$, and ${\rm ST}^{-}_{K2\pi}$ samples. Backgrounds are treated by subtracting the NLL values obtained from the $B$ sample~\cite{bkg_nll}. 

Decay amplitudes are constructed using the helicity formalism and modeled as two-body decays~\cite{LS}: $\ee\to \pi^+ R_{D^0D^-}$, $\ee\to D^- R_{D^0\pi_d^+}$, and $\ee\to D^0 R_{D^-\pi_d^+}$, where $R_{D^0D^-}$, $R_{D^0\pi_d^+}$, and $R_{D^-\pi_d^+}$ denote intermediate isobars. Parity conservation restricts the quantum numbers of these intermediate states to be either ${\rm odd}$ or ${\rm even}$. 
Assuming isospin symmetry, the amplitudes for $\ee \to D^- R_{D^0\pi_d^+}$ and $\ee \to D^0 R_{D^-\pi_d^+}$ have equal magnitudes and a relative phase of $\pi$ when $R_{D^0\pi_d^+}=D_{2}^{*}(2460)^{+}$ and $R_{D^-\pi_d^+}=\bar{D}_{2}^{*}(2460)^{0}$.  
The same relation holds for the pairs $[D^{*+}, \bar{D}^{*0}]$, $[D_{1}^{*}(2600)^{+}, \bar{D}_{1}^{*}(2600)^{0}]$, $[D_{3}^{*}(2750)^{+}, \bar{D}_{3}^{*}(2750)^{0}]$, and $[D_{1}^{*}(2760)^{+}, \bar{D}_{1}^{*}(2760)^{0}]$.

In the $R_{D^0D^-}$ system, a $J^P = 1^-$ state near the $D^0D^-$ mass threshold, denoted as $X(3790)^-$, is considered. The $X(3790)^-$ is described with a Flatté formula~\cite{flatte}, $R(m)=\frac{1}{m^{2}-m_{0}^{2}+ig m_{0}\frac{q}{m}}$, where $m_{0}$ is the nominal mass of the $X(3790)^{-}$ and $g$ is the coupling strength. The PHSP factor $q$ is defined as
\begin{widetext}
\begin{equation}
	q=\left\{\begin{array}{ll}
		\,\,\frac{\sqrt{\left(m^{2}-\left(m_{1}+m_{2}\right)^{2}\right)\left(m^{2}-\left(m_{1}-m_{2}\right)^{2}\right)}}{2 m} & {\rm above}\ D\bar{D}\ {\rm threshold}, \\ [1.6mm]
		i\frac{\sqrt{\left(\left(m_{1}+m_{2}\right)^{2}-m^{2}\right)\left(m^{2}-\left(m_{1}-m_{2}\right)^{2}\right)}}{2 m} & {\rm below}\ D\bar{D}\ {\rm threshold}.
	\end{array}\right.
\end{equation}
\end{widetext}
Since the $X(3790)^{-}$ is only considered in the $D^0D^-$ channel, $(m_{1}, m_{2})$ is taken to be $(m_{D^{0}},m_{D^{-}})$. 

For the $R_{D^0\pi_d^+}$ system, contributions from the $D^{*+}$, $D_2^*(2460)^+$, $D_1^*(2600)^+$, $D_1^*(2760)^+$, and $D_3^*(2750)^+$ are included, while the corresponding neutral states, $\bar{D}^{*0}$, $\bar{D}_2^*(2460)^0$, $\bar{D}_1^*(2600)^0$, $\bar{D}_1^*(2760)^0$, and $\bar{D}_3^*(2750)^0$, are included in the $R_{D^-\pi_d^+}$ system. 
The $D_1^*(2600)$, $D_1^*(2760)$, and $D_3^*(2750)$ states (including both the charged and neutral states, unless otherwise noted) are modeled with Breit-Wigner functions with mass-dependent widths~\cite{D2_LHCb}. 
The parameterizations of the $D^{*}$ mesons follow Ref.~\cite{D2_LHCb}, while the $D_2^*(2460)$ is modeled assuming $D\pi$ and $D^*\pi$ are the dominant decays, as detailed in the Supplemental Material~\cite{supp}. 
In the PWA fit, the mass and $D^0D^-$ coupling strength of the $X(3790)^-$, along with the masses and widths of the $D_2^*(2460)$, $D_1^*(2600)$, and $D_1^*(2760)$, are left floating, and the masses and widths of $D_1^*(2600)^+$ and $\bar{D}_1^*(2600)^0$ are constrained to be identical, as well as those of the $D_1^*(2760)^+$ and $\bar{D}_1^*(2760)^0$. The mass and width of the $D_3^*(2750)$ are fixed to their nominal values from the Particle Data Group (PDG)~\cite{pdg}.

Figure~\ref{pic:pwa_base} shows comparisons between the invariant mass distributions from data and the PWA fit projections with and without the inclusion of the $X(3790)^-$ component. 
The fit quality is evaluated using the $\chi^2/n_{\rm bin}$ metric, where $n_{\rm bin}$ is the number of bins and $\chi^2 = \sum_{i=1}^{n_{\rm bin}} \frac{(n^S_i - n^B_i - v_i)^2}{n^S_i + n^B_i}$, with $n^S_i$, $n^B_i$, and $v_i$ denoting the number of events in the $i^{\rm th}$ bin from the $S$ sample, $B$ sample, and PWA fit, respectively. With the inclusion of the $X(3790)^-$ component, the values of $\chi^{2}/n_{\rm bin}$ are 1.38 (a), 0.99 (b), 1.09 (c), 0.99 (d), 0.97 (e), and 1.13 (f).
The data samples are well described by the PWA fit that includes the $X(3790)^-$ component, which is taken as the nominal fit. The extracted parameters are shown in Table~\ref{tab:pwa-res-param}, together with the significances of the $D_1^*(2600)$, $D_1^*(2760)$, and $D_3^*(2750)$ components, derived from the changes in the NLL and the number of degrees of freedom, and with the systematic effects considered as described below.
The significance of the $X(3790)^-$ signal is evaluated to be $7.7\sigma$ from the changes in the NLL and the number of degrees of freedom.  
The look-elsewhere effect~\cite{lee} is accounted for using 320 pseudo-experiments generated under the null-$X(3790)^-$ hypothesis.  
In each pseudo-experiment, the global maximum of $-2\ln({\cal L}_1/{\cal L}_0)$, where ${\cal L}_0$ and ${\cal L}_1$ are the NLL values for the null and alternative hypotheses, is found by scanning the $X(3790)^-$ mass and its $D^0D^-$ coupling strength over the kinematically allowed region, rather than constraining them to the data-fit values~\cite{Zcs_3985,Pcs_4459}.  
The distribution of $-2\ln({\cal L}_1/{\cal L}_0)$ is well described by a $\chi^2$ distribution with 20.0 degrees of freedom.  
After inclusion of systematic uncertainties described below, with an observed value of $-2\ln({\cal L}_1/{\cal L}_0)=71.5$, we obtain a significance of $5.3\sigma$.

\begin{table*}[htbp]
	\centering
	\caption{Masses ($m_0$), widths ($\Gamma_0$)/$g$, fit fractions (FF), and significances ($\mathcal{S}$) of the resonances. The first uncertainties are statistical and the second systematic. }
	\begin{tabular}{lcccc}
		\hline\hline\noalign{\smallskip}
		Resonance & $m_0$ (MeV/$c^2$) & $\Gamma_0$/$g$ & FF (\%) & $\mathcal{S}$ \\
		\noalign{\smallskip}\hline\noalign{\smallskip}
	$D^{*}_{2}(2460)^{+}$  & $2469.8\pm 0.9\pm 2.4$ & $(42.6\pm 1.6\pm 0.7)$ MeV & $9.1\pm0.9$ & - \\
	$\bar{D}^{*}_{2}(2460)^{0}$ & $2468.4\pm 0.9\pm 1.8$ & $(43.1\pm 1.5\pm 2.8)$ MeV & $9.0\pm0.9$ &-\\
	$D^{*}_{1}(2760)$  & $2716.8\pm 2.4\pm 6.3$ & $(91\pm 5\pm 18)$ MeV & $2.3\pm0.3$ & $6.4\sigma$ \\
	$D^{*}_{1}(2600)$  & $2563\pm 14\pm 22$ & $(222\pm 25\pm 38)$ MeV & $3.6\pm0.4$ & $5.5\sigma$ \\
	$X(3790)^{-}$      & $3794.7\pm 6.0\pm 8.7$ & $(1.6\pm 0.2\pm 0.2)$ GeV & $1.3\pm0.3$ & $5.3\sigma$ \\
	$D^{*}_{3}(2750)$  & - & - & $2.9\pm 0.1$ & $11\sigma$ \\
	$D^{*}$  & - & - & $78.6\pm 2.2$ & - \\
		\noalign{\smallskip}\hline\hline
	\end{tabular}
	\label{tab:pwa-res-param}
\end{table*}

Several cross checks are performed to validate the existence of the $X(3790)^-$. 
Fits assigning alternative $J^P$ values ($J<4$) to the $X(3790)^-$ give worse descriptions of the data, with increases in NLL by at least 19 units. 
To account for possible PHSP contributions, a constant amplitude with a complex phase is added to the nominal amplitude set.
This yields an NLL decrease of 5 units, corresponding to a significance of $2.7\sigma$, calculated from the changes in NLL and the number of degrees of freedom (the same method is used throughout this paragraph). When this term is included, the significance of the $X(3790)^-$ is $8.9\sigma$.
To assess potential non-resonant contributions, individual amplitudes with $J^P = 1^-$, $2^+$, or $3^-$ assigned to the $D^0D^-$ or $D^-\pi_d^+$ systems are sequentially added to the nominal amplitude set, setting $R_{D^0D^-}(m)$ or $R_{D^-\pi_d^+}(m)=1$ in each case.  
The $D^0D^-$ system with $J^P=2^+$ produces the largest effect, decreasing the NLL by 4 units, corresponding to a significance of $2.2\sigma$.  When this component is included, the significance of the $X(3790)^-$ remains $8.2\sigma$.
The parameters of the $D_{2}^{*}(2460)$, $D_{1}^{*}(2600)$, $D_{1}^{*}(2760)$, and $D_{3}^{*}(2750)$ are fixed to their nominal values from the PDG~\cite{pdg}, and when the $X(3790)^-$ is included with floating parameters, the NLL increases by 76 units, and its significance remains over $10\sigma$ (the data and PWA projections are shown in the Supplemental Material~\cite{supp}).

The systematic uncertainties on the $X(3790)^-$, $D_{2}^*(2460)$, $D_1^*(2600)$, and $D_1^*(2760)$ resonance parameters arise from $D_{3}^*(2750)$ resonance parameters (\romannumeral1), the parameterization of the $D_2^*(2460)$ (\romannumeral2), the radius parameter when constructing helicity amplitude (\romannumeral3), background modeling (\romannumeral4), width of the $\bar{D}^{*0}$ (\romannumeral5), ISR effect (\romannumeral6), and detector resolution (\romannumeral7); whereas those on the significances of the $X(3790)^-$, $D_1^*(2600)$, $D_1^*(2760)$, and $D_3^*(2750)$ arise from (\romannumeral1) to (\romannumeral6).

In the nominal fit, the mass and width of the $D_3^*(2750)$ are fixed to the PDG values~\cite{pdg}, while in an alternative fit they are allowed to vary within their uncertainties. 
The width of the $D_2^*(2460)$ is modeled using the ratio of $\mathcal{B}(D_2^*(2460)\to D\pi)/\mathcal{B}(D_2^*(2460)\to D^*\pi)$ reported by the PDG~\cite{pdg}. The prediction from Ref.~\cite{HQL_D} is adopted, and alternative fits are performed to assess the associated systematic uncertainty.
The uncertainty from the radius parameter used in the amplitude construction is evaluated following Ref.~\cite{sys_radius}. 
Systematic uncertainties from background modeling and detector resolution are described in the Supplemental Material~\cite{supp}.
The $\bar{D}^{*0}$ width is fixed to its PDG upper limit~\cite{pdg} in the nominal fit, while the $D^{*+}$ width is used in alternative fits. 
The ISR correction for the $\ee\to \DDpi$ process is based on a flat cross section line shape in the nominal MC generation, while the one measured by Belle~\cite{DDPI_Belle} is used as an alternative.

The total systematic uncertainties, as summarized in the Supplemental Material~\cite{supp}, are obtained by summing all individual sources in quadrature under the assumption that they are uncorrelated.

In summary, we perform a PWA on the $\ee\to \DDpi$ process with the data sample taken at $\sqrt{s}=4682$~MeV. 
A charged charmoniumlike structure with $J^{P}=1^{-}$ around the $D^{0}D^{-}$ threshold is observed for the first time with a mass of $(3794.7\pm 6.0\pm 8.7)$~MeV/$c^{2}$ and a coupling strength to $D^0D^-$ of $(1.6 \pm 0.2 \pm 0.2)$~GeV, where the first uncertainties are statistical and the second systematic. 
The significance of this state is 5.3$\sigma$ including systematic uncertainties and the look-elsewhere effect. 
The masses and widths of the $D^{*}_{2}(2460)^{+}$, $\bar{D}^{*}_{2}(2460)^{0}$, $D^{*}_{1}(2600)$, and $D^{*}_{1}(2760)$ are also measured as shown in Table~\ref{tab:pwa-res-param}.
The observation of this charged structure near the $D^0D^-$ threshold, the $X(3790)^-$, offers new insights into the nature of the $X(3872)$ and other charmoniumlike states. In addition, the updated parameters of the excited charm mesons contribute to a better understanding of the strong interaction between a charmed quark and a light antiquark.

\acknowledgments

The BESIII Collaboration thanks the staff of BEPCII (https://cstr.cn/31109.02.BEPC) and the IHEP computing center for their strong support. This work is supported in part by National Key R\&D Program of China under Contracts Nos. 2023YFA1606000, 2023YFA1606704, 2025YFA1613900; National Natural Science Foundation of China (NSFC) under Contracts Nos. 11635010, 11935015, 11935016, 11935018, 12025502, 12035009, 12035013, 12061131003, 12192260, 12192261, 12192262, 12192263, 12192264, 12192265, 12221005, 12225509, 12235017, 12342502, 12361141819; the Chinese Academy of Sciences (CAS) Large-Scale Scientific Facility Program; the Strategic Priority Research Program of Chinese Academy of Sciences under Contract No. XDA0480600; CAS under Contract No. YSBR-101; 100 Talents Program of CAS; The Institute of Nuclear and Particle Physics (INPAC) and Shanghai Key Laboratory for Particle Physics and Cosmology; ERC under Contract No. 758462; German Research Foundation DFG under Contract No. FOR5327; Istituto Nazionale di Fisica Nucleare, Italy; Knut and Alice Wallenberg Foundation under Contracts Nos. 2021.0174, 2021.0299, 2023.0315; Ministry of Development of Turkey under Contract No. DPT2006K-120470; National Research Foundation of Korea under Contract No. NRF-2022R1A2C1092335; National Science and Technology fund of Mongolia; Polish National Science Centre under Contract No. 2024/53/B/ST2/00975; STFC (United Kingdom); Swedish Research Council under Contract No. 2019.04595; U. S. Department of Energy under Contract No. DE-FG02-05ER41374.

\end{document}



\begin{center}
   {{\bf \boldmath Supplemental Material for ``Observation of a Charged Charmoniumlike Structure in $\ee\to\DDpi+c.c.$"}}
\end{center}
\author{
M.~Ablikim$^{1}$\BESIIIorcid{0000-0002-3935-619X},
M.~N.~Achasov$^{4,c}$\BESIIIorcid{0000-0002-9400-8622},
P.~Adlarson$^{83}$\BESIIIorcid{0000-0001-6280-3851},
X.~C.~Ai$^{89}$\BESIIIorcid{0000-0003-3856-2415},
C.~S.~Akondi$^{31A,31B}$\BESIIIorcid{0000-0001-6303-5217},
R.~Aliberti$^{39}$\BESIIIorcid{0000-0003-3500-4012},
A.~Amoroso$^{82A,82C}$\BESIIIorcid{0000-0002-3095-8610},
Q.~An$^{79,65,\dagger}$,
Y.~H.~An$^{89}$\BESIIIorcid{0009-0008-3419-0849},
Y.~Bai$^{63}$\BESIIIorcid{0000-0001-6593-5665}
% ...
\\
\vspace{0.2cm}
(BESIII Collaboration)\\
\vspace{0.2cm} {\it
$^{1}$ Institute of High Energy Physics, Beijing 100049, People's Republic of China\\
% ...
$^{4}$ Budker Institute of Nuclear Physics SB RAS (BINP), Novosibirsk 630090, Russia\\
% ...
$^{31}$ INFN Sezione di Ferrara, (A)INFN Sezione di Ferrara, I-44122, Ferrara, Italy; (B)University of Ferrara, I-44122, Ferrara, Italy\\
% ...
$^{39}$ Johannes Gutenberg University of Mainz, Johann-Joachim-Becher-Weg 45, D-55099 Mainz, Germany\\
% ...
$^{63}$ Southeast University, Nanjing 211100, People's Republic of China\\
% ...
$^{65}$ State Key Laboratory of Particle Detection and Electronics, Beijing 100049, Hefei 230026, People's Republic of China\\
% ...
$^{79}$ University of Science and Technology of China, Hefei 230026, People's Republic of China\\
% ...
$^{82}$ University of Turin and INFN, (A)University of Turin, I-10125, Turin, Italy; (B)University of Eastern Piedmont, I-15121, Alessandria, Italy; (C)INFN, I-10125, Turin, Italy\\
$^{83}$ Uppsala University, Box 516, SE-75120 Uppsala, Sweden\\
% ...
$^{89}$ Zhengzhou University, Zhengzhou 450001, People's Republic of China\\
\vspace{0.2cm}
$^{\dagger}$ Deceased\\
% ...
$^{c}$ Also at the Novosibirsk State University, Novosibirsk, 630090, Russia\\
% ...
}}
\maketitle
\bigskip

\section{The $S$ and $B$ Samples}

\begin{figure*}[!htbp]
    \centering
    \includegraphics[width=0.45\textwidth]{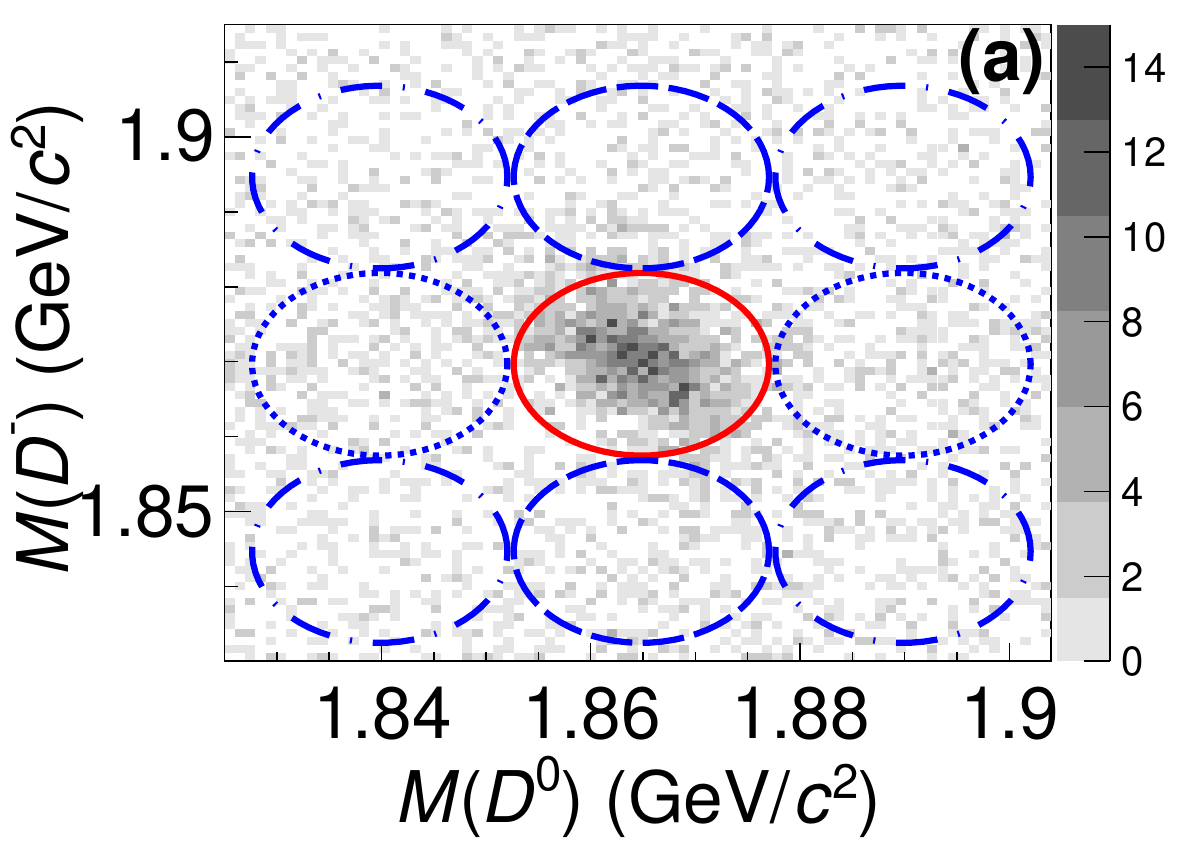}
    \includegraphics[width=0.45\textwidth]{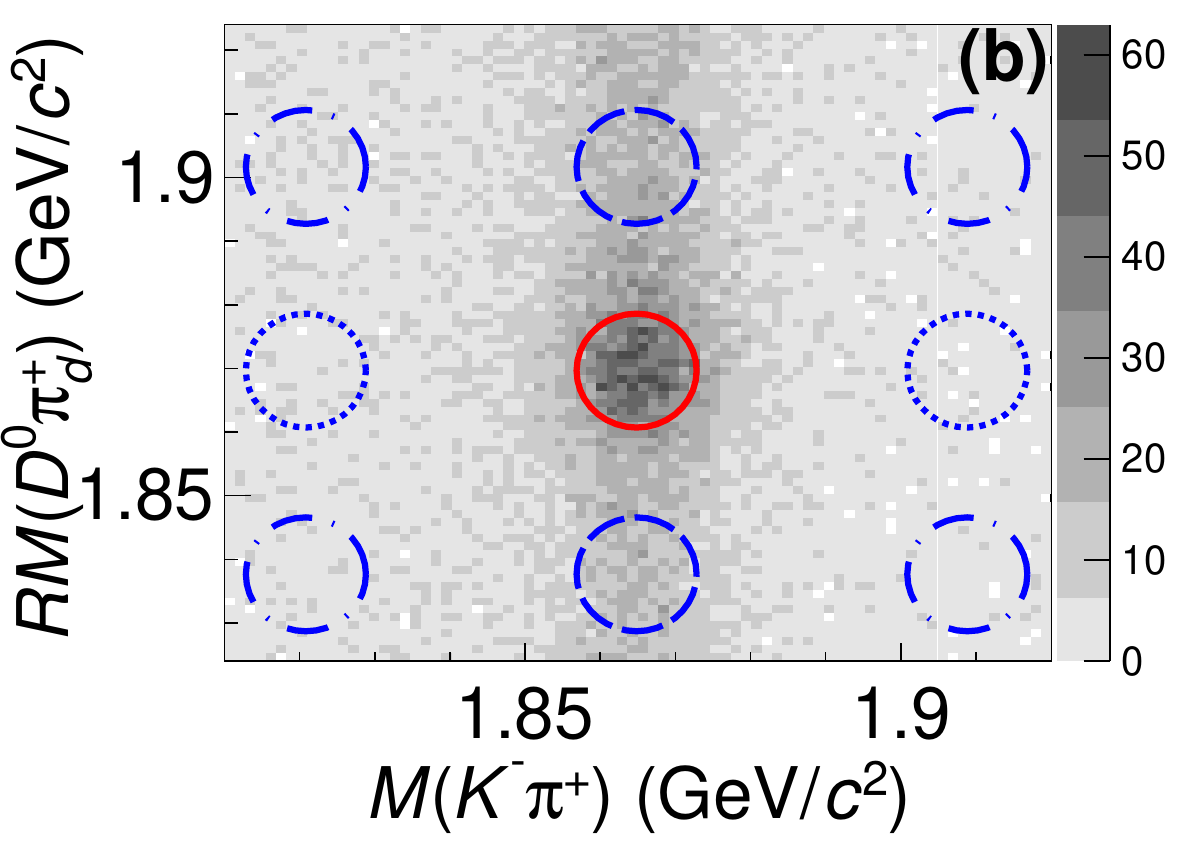}
    \includegraphics[width=0.45\textwidth]{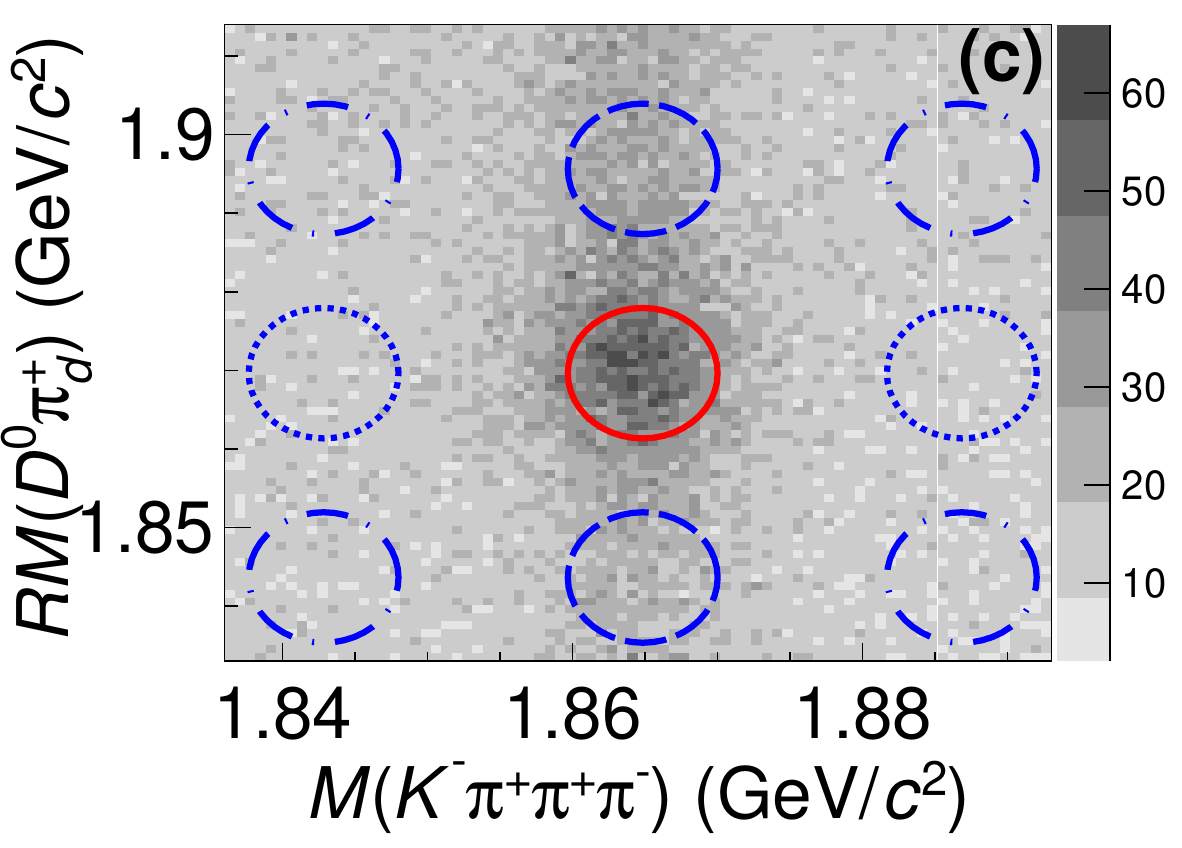}
    \includegraphics[width=0.45\textwidth]{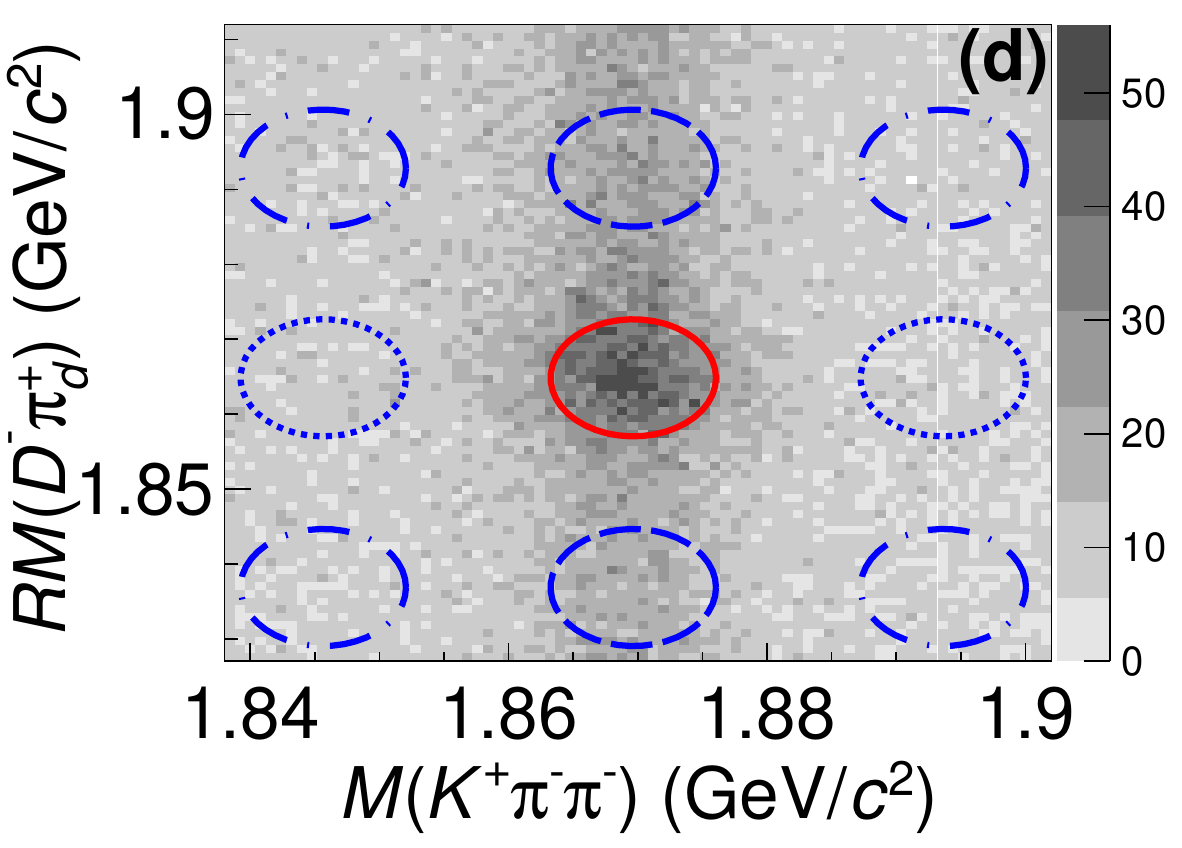}
    \caption{The scatter plots of $M(D^0)$ versus $M(D^-)$ from the data in ${\rm DT}$ sample (a); $M(K^-\pi^+)$ versus $RM(D^0\pi_d^+)$ from the data in ${\rm ST}^{0}_{K\pi}$ sample (b); $M(K^-\pi^+\pi^+\pi^-)$ versus $RM(D^0\pi_d^+)$ from the data in ${\rm ST}^{0}_{K3\pi}$ sample (c); and $M(K^+\pi^-\pi^-)$ versus $RM(D^-\pi_d^+)$ from the data in ${\rm ST}^{-}_{K2\pi}$ sample (d). The red solid, blue dashed, blue dotted, and blue dash-dotted ellipses represent events in $S$, $B_1$, $B_2$, and $B_3$ samples, respectively.}
    \label{pic:mass}
\end{figure*}

As shown in Fig.~\ref{pic:mass}, the signal and sideband regions for $D^0$ and $D^-$ candidates in both ${\rm ST}$ and ${\rm DT}$ samples are defined with elliptical selections. 
The ellipses are defined as
\begin{equation}
    \label{ellip}
	\sqrt{\left(\frac{M_1 - {\rm mean}_1}{\sigma_{M_1}}\right)^2 + 
              \left(\frac{M_2 - {\rm mean}_2}{\sigma_{M_2}}\right)^2}<\delta.
\end{equation}

For ${\rm ST}^{0}_{K\pi}$ sample, $M_1$ represents the invariant mass of $K^-\pi^+$ [$M(K^-\pi^+)$], and $M_2$ represents the recoiling mass of $D^0\pi_d^+$ with the invariant mass of $K^-\pi^+$ constrained to $m_{D^0}$ [$RM(D^0\pi_d^+)$]. 
Here, $\sigma_{M_1}$ and $\sigma_{M_2}$ represent the mass resolutions of $M(K^-\pi^+)$ and $RM(D^0\pi_d^+)$, 
and they are obtained by the fits on the corresponding distributions with a Gaussian function as signal shape and a second order Chybeshev polynomial function as background shape, the standard deviations of the Gaussian function are taken as the mass resolution, which are $\sigma_{M_1}=6.7$~MeV/$c^{2}$ and $\sigma_{M_2}=7.5$~MeV/$c^{2}$.
The estimations of $\sigma_{M_1}$ and $\sigma_{M_2}$ for the other samples follow the same way.

For ${\rm ST}^{0}_{K3\pi}$ sample, $M_1$ represents the invariant mass of $K^-\pi^+\pi^+\pi^-$ [$M(K^-\pi^+\pi^+\pi^-)$], $M_2$ represents the recoiling mass of $D^0\pi_d^+$ with the invariant mass of $K^-\pi^+\pi^+\pi^-$ constrained to $m_{D^0}$ [$RM(D^0\pi_d^+)$], and $\sigma_{M_1}=4.2$~MeV/$c^{2}$ and $\sigma_{M_2}=6.9$~MeV/$c^{2}$.
For ${\rm ST}^{-}_{K2\pi}$ sample, $M_1$ represents the invariant mass of $K^+\pi^-\pi^-$ [$M(K^+\pi^-\pi^-)$], $M_2$ represents the recoiling mass of $D^-\pi_d^+$ with the invariant mass of $K^+\pi^-\pi^-$ constrained to $m_{D^-}$ [$RM(D^-\pi_d^+)$], and $\sigma_{M_1}=5.3$~MeV/$c^{2}$ and $\sigma_{M_2}=6.5$~MeV/$c^{2}$.
For ${\rm DT}$ sample, $M_1$ and $M_2$ represent the reconstructed $D^0$ and $D^-$ invariant masses [$M(D^0)$ and $M(D^-)$], and $\sigma_{M_1}=5.1$~MeV/$c^{2}$ and $\sigma_{M_2}=5.1$~MeV/$c^{2}$.

In Eq.~(\ref{ellip}), $\delta$ defines the size of the ellipse, which is 2.0 in ${\rm DT}$ sample and 1.2 in ${\rm ST}$ samples, and $({\rm mean}_1, \, {\rm mean}_2)$ represents the center of the ellipses.
For the signal region, $({\rm mean}_1, \, {\rm mean}_2)$ is defined as $(m_{D^0}, \, m_{D^-})$ in ${\rm DT}$ and ${\rm ST}^{0}_{K\pi, K3\pi}$ samples or $(m_{D^-}, \, m_{D^0})$ in ${\rm ST}^{-}_{K2\pi}$ sample ($S$ sample). 
The sideband regions are also defined based on $({\rm mean}_1, \, {\rm mean}_2)$, which are $(m_{D^0}, \, m_{D^-}\pm 2 d_{D^-})$ ($B_1$ sample), $(m_{D^0}\pm 2 d_{D^0}, \, m_{D^-})$ ($B_2$ sample),
and $(m_{D^0}\pm 2 d_{D^0}, \, m_{D^-}\pm 2 d_{D^-})$ ($B_3$ sample) in ${\rm DT}$ and ${{\rm ST}^{0}_{K\pi, K3\pi}}$ samples, or $(m_{D^-}, \, m_{D^0}\pm 2 d_{D^0})$ ($B_1$ sample), $(m_{D^-}\pm 2 d_{D^-}, \, m_{D^0})$ ($B_2$ sample), and $(m_{D^-}\pm 2 d_{D^-}, \, m_{D^0}\pm 2 d_{D^0})$ ($B_3$ sample) in ${{\rm ST}^{-}_{K2\pi}}$ sample.
For ${\rm ST}^{0}_{K\pi}$ sample, $d_{D^0}=22$~MeV/$c^{2}$ and $d_{D^-}=16$~MeV/$c^{2}$;
for ${\rm ST}^{0}_{K3\pi}$ sample, $d_{D^0}=11$~MeV/$c^{2}$ and $d_{D^-}=13$~MeV/$c^{2}$;
for ${\rm DT}$ sample, $d_{D^0}=16$~MeV/$c^{2}$ and $d_{D^-}=14$~MeV/$c^{2}$;
and for ${\rm ST}^{-}_{K2\pi}$ sample, $d_{D^0}=14$~MeV/$c^{2}$ and $d_{D^-}=12$~MeV/$c^{2}$.
The $B$ sample is defined as 
\begin{equation}
    B = f_1\cdot B_1+f_2\cdot B_2 - f_3\cdot B_3,
\end{equation}
where $f_1=0.5$, $f_2=0.5$, and $f_3=0.25$ are the normalization factors assuming a linear mass dependence in the background distributions.

\section{Dynamic Part}

For the $D^{*}_{2}(2460)$, a coupled channel relativistic Breit-Wigner formula is used,
\begin{equation}
	R(m)=\frac{1}{m^2-m_0^2+im_0 \Gamma_{\rm tot}(m)},
\end{equation}
where $m_{0}$ is the nominal mass for $D^{*}_{2}(2460)$ and the mass dependent width is
\begin{equation}
	\Gamma_{\rm tot}(m)=\sum_{a}\Gamma_{a}(m),
\end{equation}
\begin{equation}
	\Gamma_{a}(m)=R_{a}\Gamma_0\left(\frac{q}{q_0}\right)^{2 l_0+1} \frac{m_0}{m} {B_{l_0}^{\prime}}^{2}\left(q, q_0, d\right),
\end{equation}
where $\Gamma_0$ is the nominal width for $D^{*}_{2}(2460)^{+/0}$, $l_{0}=2$ is orbital angular momentum of $\pi$ meson decaying from $D^{*}_{2}(2460)\to D\pi$ or $D^*\pi$, $q$ is the momentum of $\pi$ meson in the rest frame of $D_2^*(2460)$, the normalization factor $q_{0}$ is calculated at the nominal mass of $D_2^*(2460)$, the factor $B_l^{\prime}\left(q, q_0, d\right)$ is the reduced Blatt-Weisskopf factor~\cite{bf}, and the radius $d$ is the same as that in Ref.~\cite{radius}.

According to the isospin symmetry, the relationships between the amplitudes for different $D_{2}^{*}(2460)$ decay channels are: 
 $2\times |\mathcal{M}_{D^{*}_{2}(2460)^{0} \to D^{0}\pi^{0}}|^{2} = |\mathcal{M}_{D^{*}_{2}(2460)^{0} \to D^{+}\pi^{-}}|^{2}$, 
 $2\times |\mathcal{M}_{D^{*}_{2}(2460)^{0} \to D^{*0}\pi^{0}}|^{2} = |\mathcal{M}_{D^{*}_{2}(2460)^{0} \to D^{*+}\pi^{-}}|^{2}$, 
 $|\mathcal{M}_{D^{*}_{2}(2460)^{+} \to D^{0}\pi^{+}}|^{2} = 2 \times |\mathcal{M}_{D^{*}_{2}(2460)^{+} \to D^{+}\pi^{0}}|^{2}$, and 
 $|\mathcal{M}_{D^{*}_{2}(2460)^{+} \to D^{*0}\pi^{+}}|^{2} = 2 \times |\mathcal{M}_{D^{*}_{2}(2460)^{+} \to D^{*+}\pi^{0}}|^{2}$. 
According to PDG \cite{pdg}, the branching ratio is $\frac{\mathcal{B}(D^{*}_{2}(2460) \to D\pi)}{\mathcal{B}(D^{*}_{2}(2460) \to D^{*}\pi)} = (1.52 \pm 0.14)$. Then the ratios between different $D^{*}_{2}(2460)$ decay channels ($R_a$) are estimated as
\begin{equation}
	R_{D_2^*(2460)^{+}} \rightarrow\left\{\begin{array}{c}
		D^{+} \pi^0 = 0.25 \\
		D^0 \pi^{+} = 0.50 \\
		D^{*+} \pi^0 = 0.08 \\
		D^{* 0} \pi^{+} = 0.17
	\end{array}\right.,
\end{equation}
\begin{equation}
	R_{D_2^*(2460)^{0}} \rightarrow\left\{\begin{array}{c}
		D^{0} \pi^0 = 0.25 \\
		D^{+} \pi^{-} = 0.50 \\
		D^{*0} \pi^0 = 0.08 \\
		D^{* +} \pi^{-} = 0.17
	\end{array}\right. .
\end{equation}

\section{PWA Results}

\begin{figure*}[!htbp]
    \centering
    \includegraphics[width=0.3\textwidth]{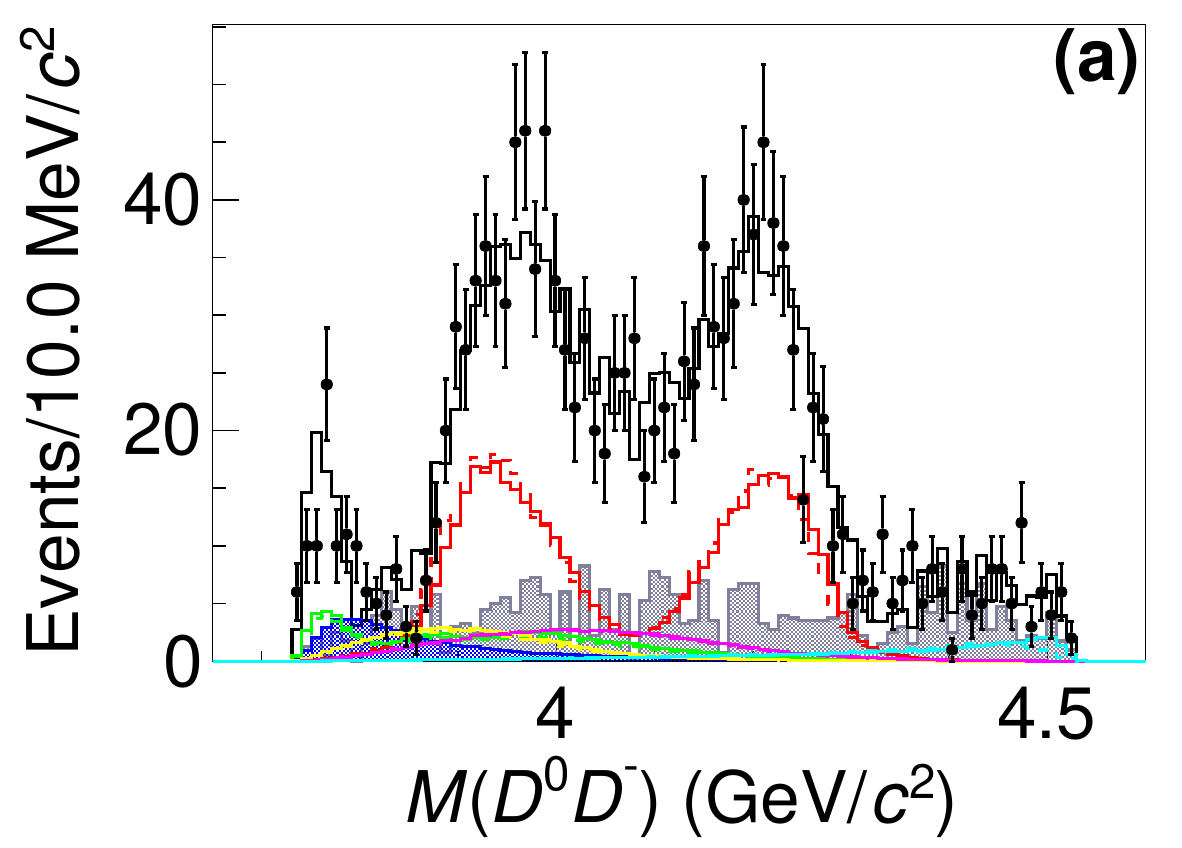}
    \includegraphics[width=0.3\textwidth]{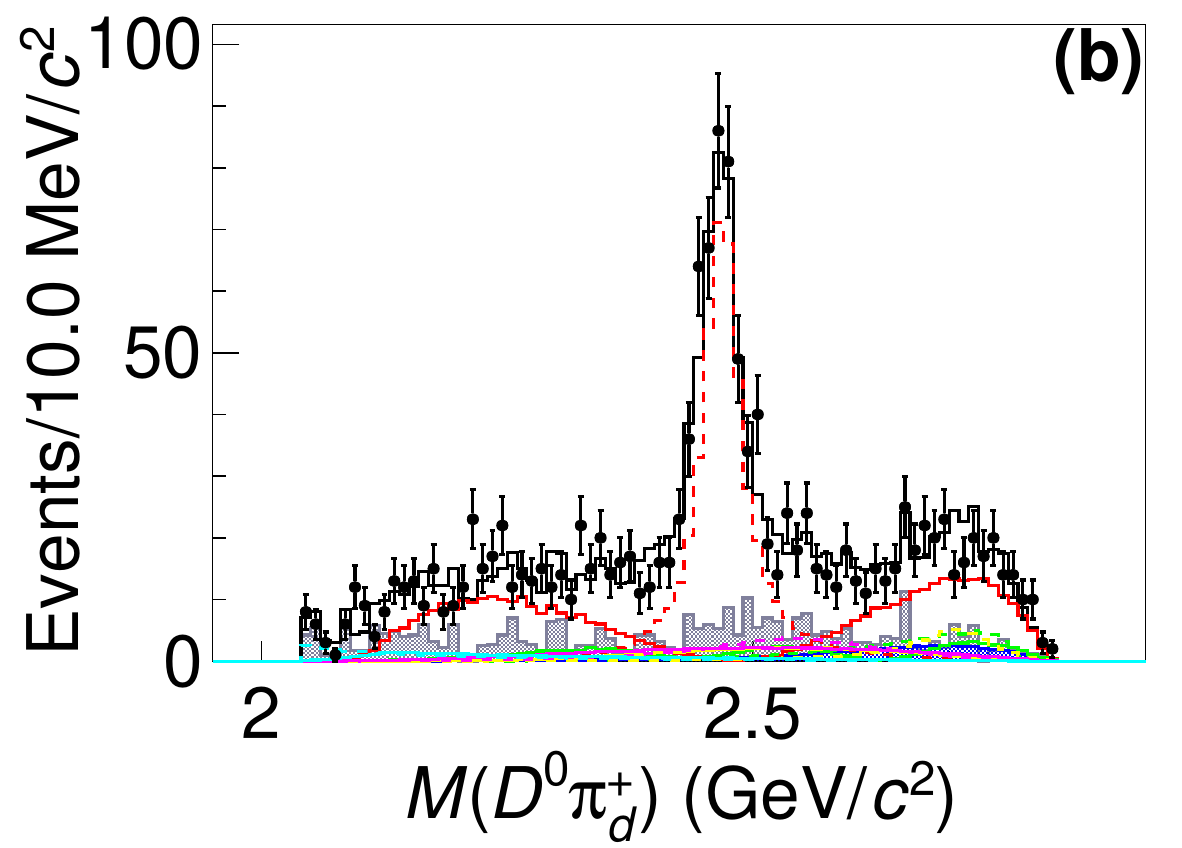}
    \includegraphics[width=0.3\textwidth]{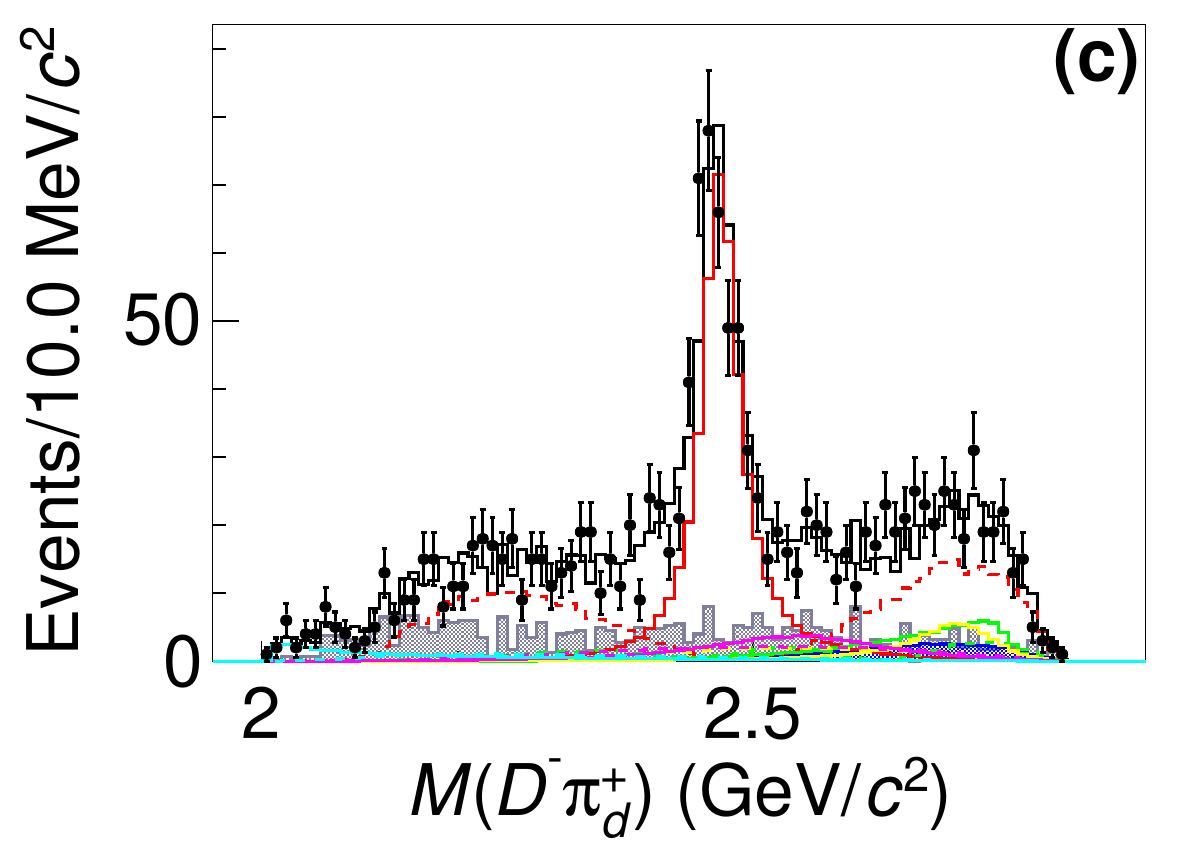}
    \includegraphics[width=0.3\textwidth]{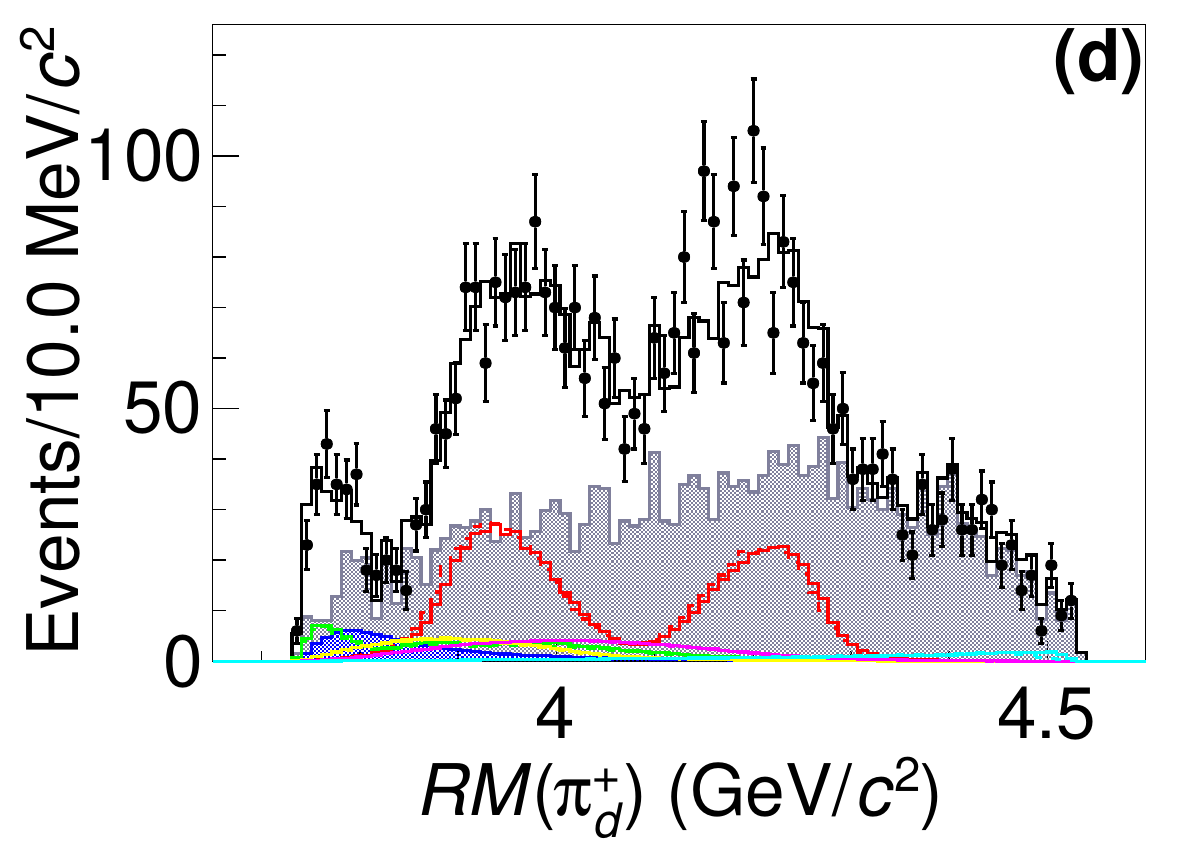}
    \includegraphics[width=0.3\textwidth]{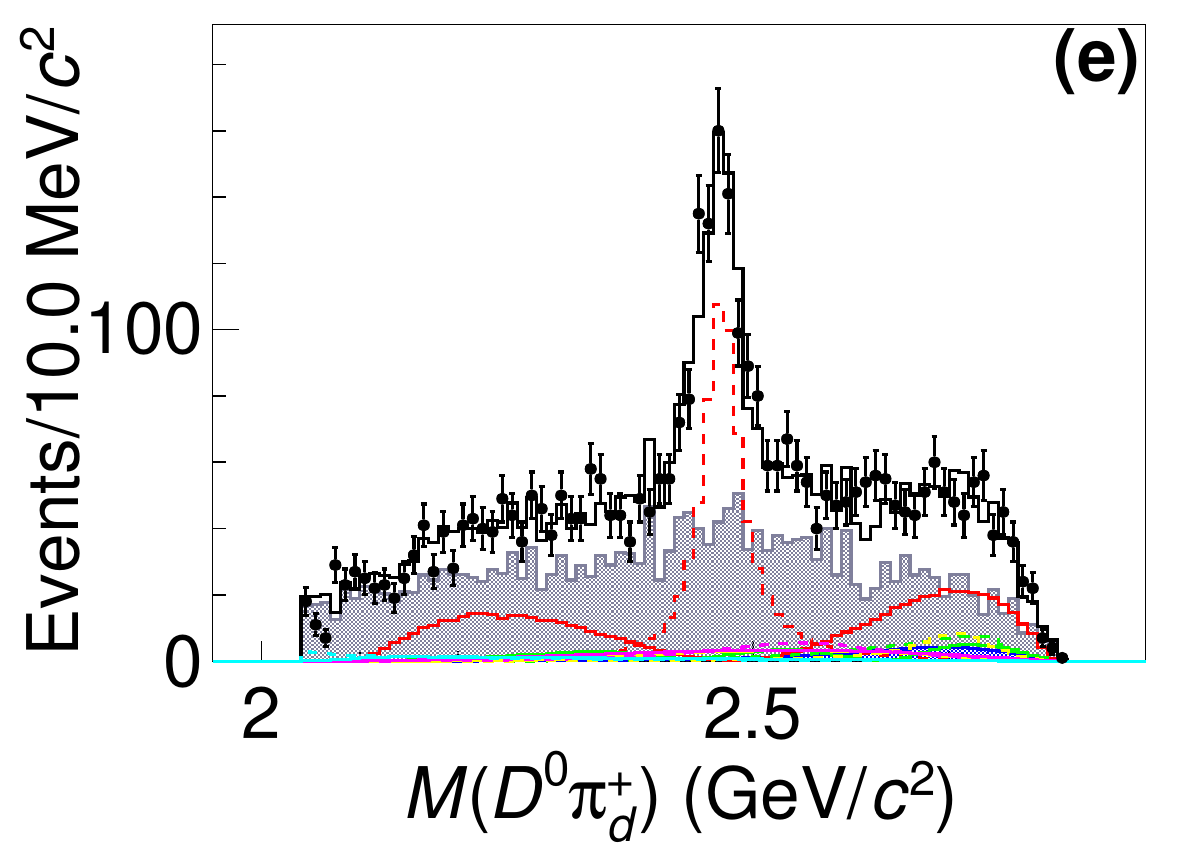}
    \includegraphics[width=0.3\textwidth]{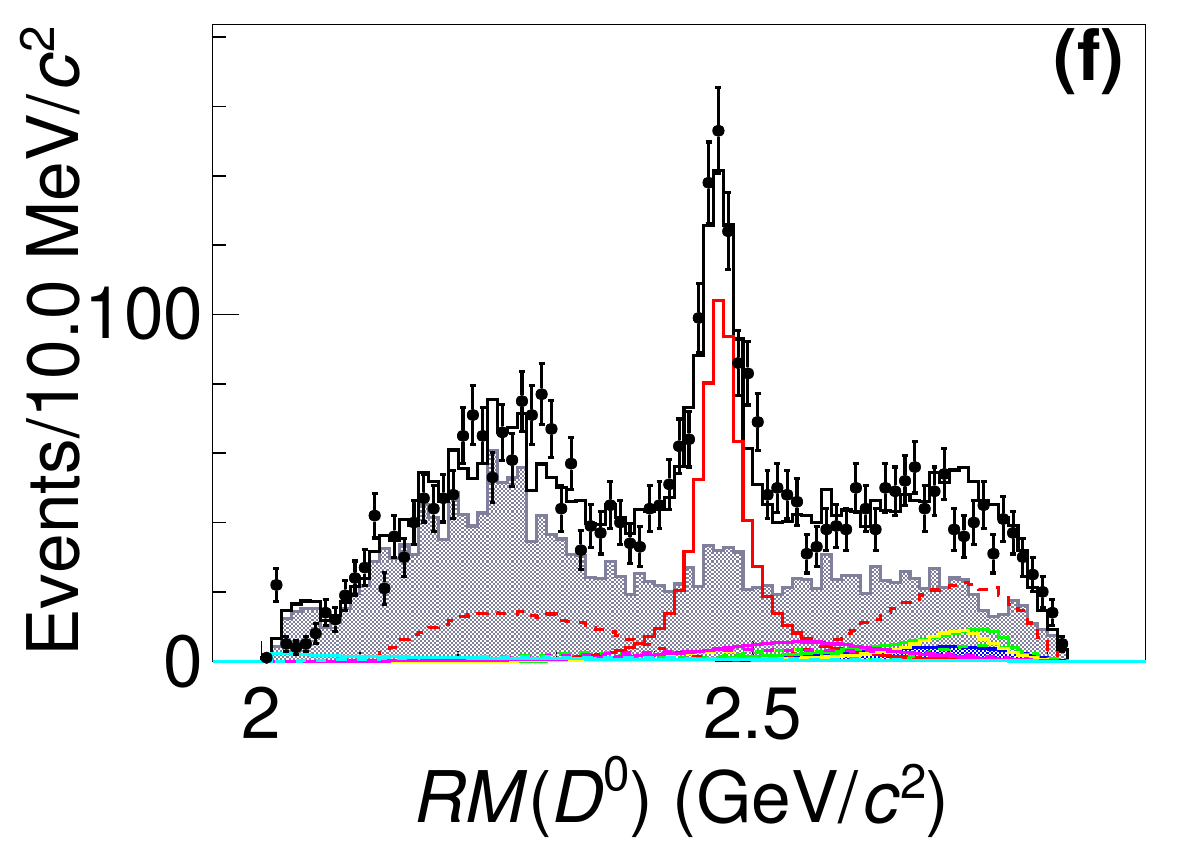}
    \includegraphics[width=0.3\textwidth]{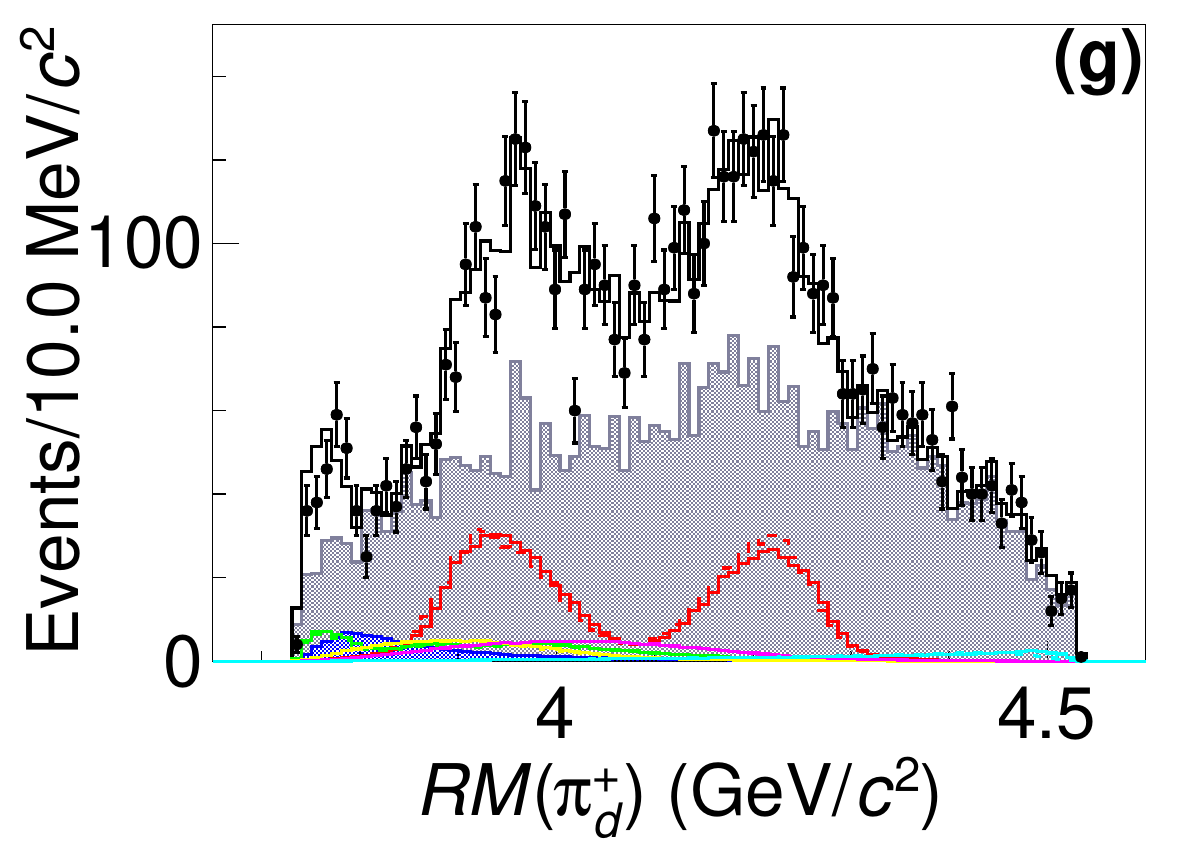}
	\includegraphics[width=0.3\textwidth]{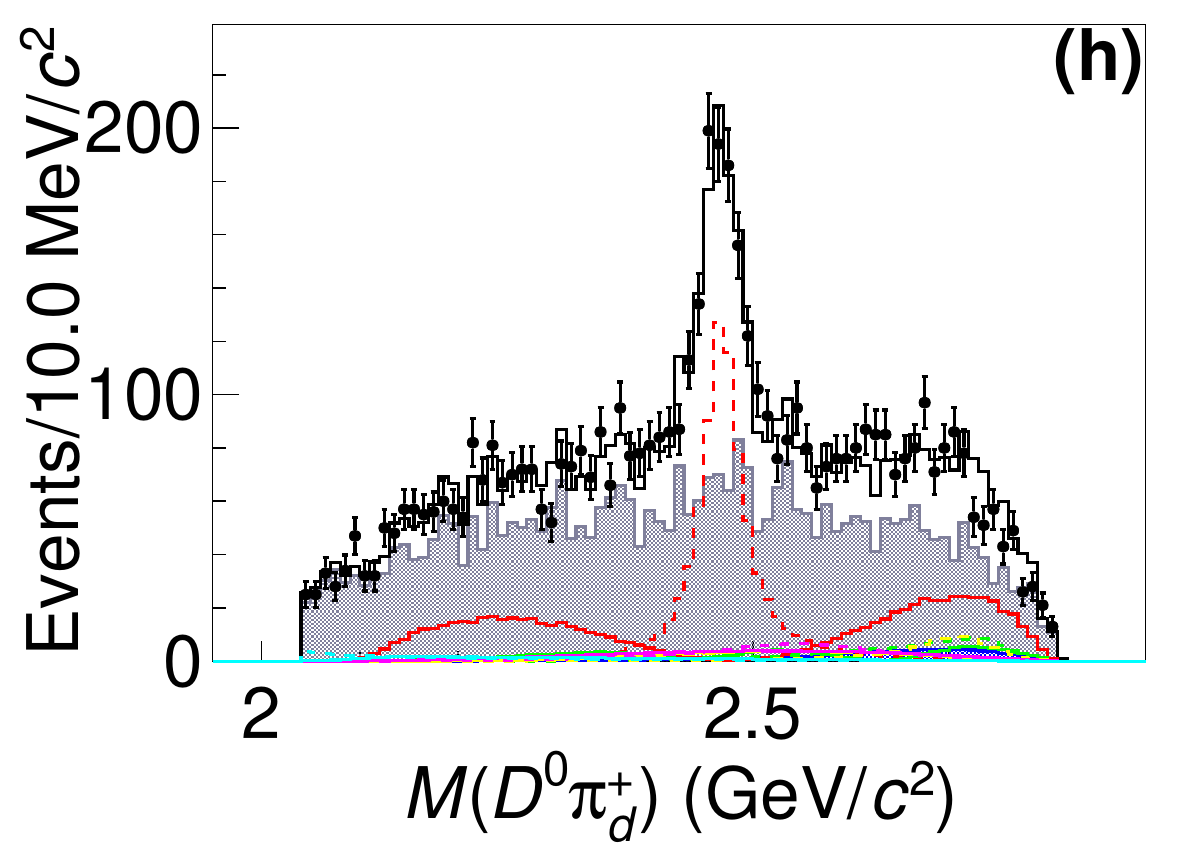}
	\includegraphics[width=0.3\textwidth]{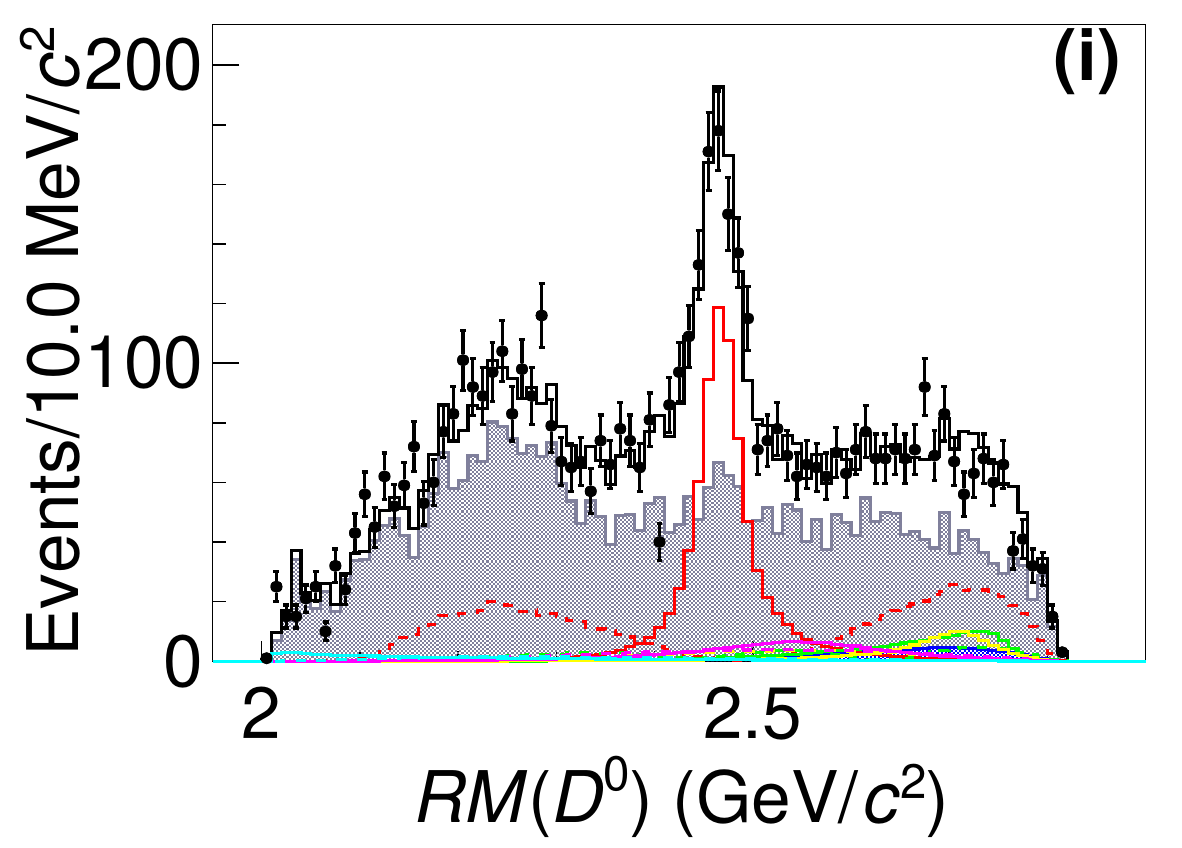}
    \includegraphics[width=0.3\textwidth]{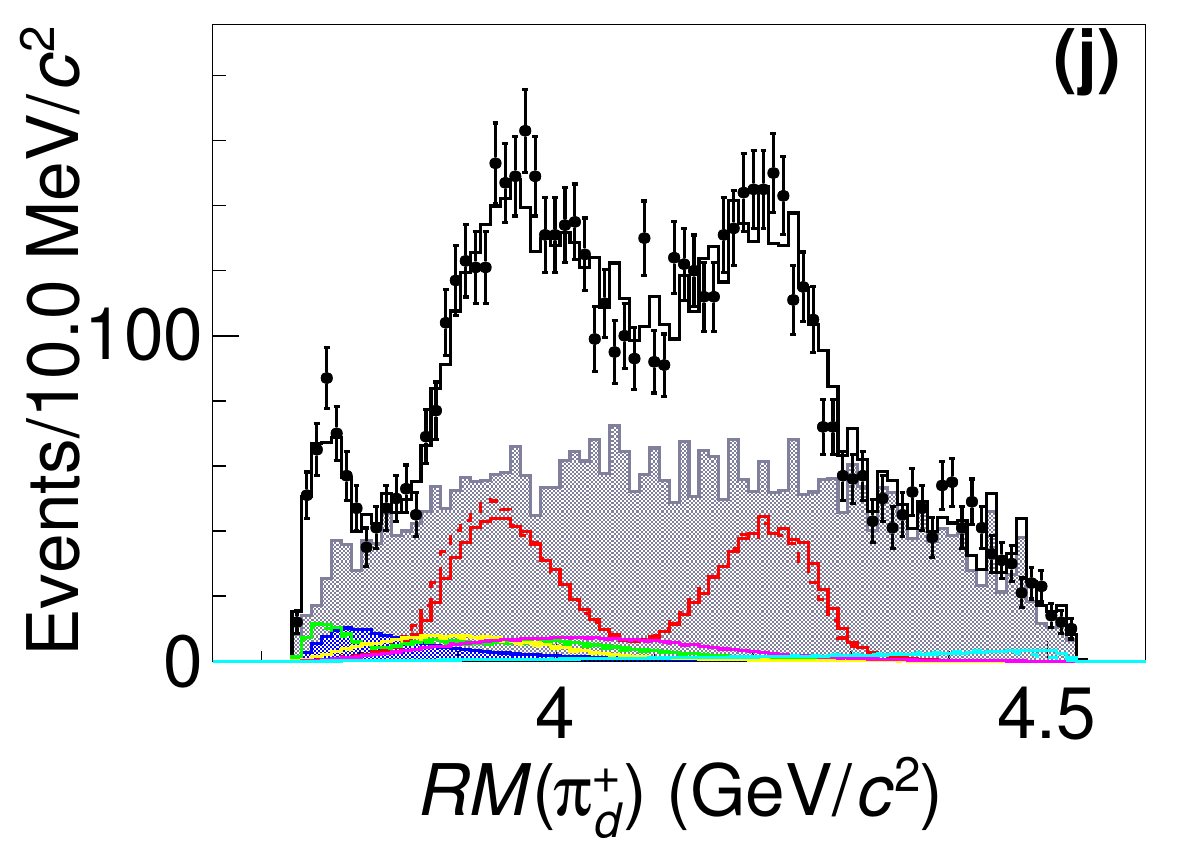}
	\includegraphics[width=0.3\textwidth]{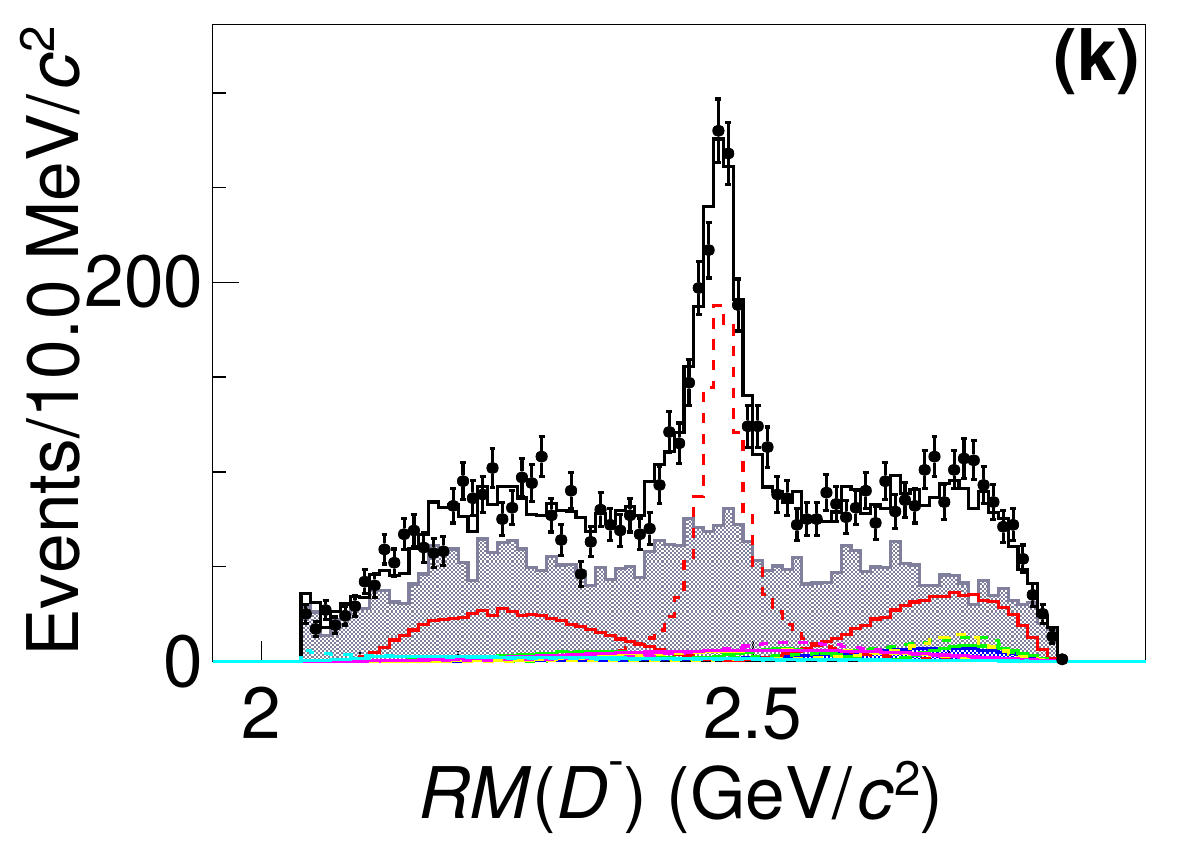}
	\includegraphics[width=0.3\textwidth]{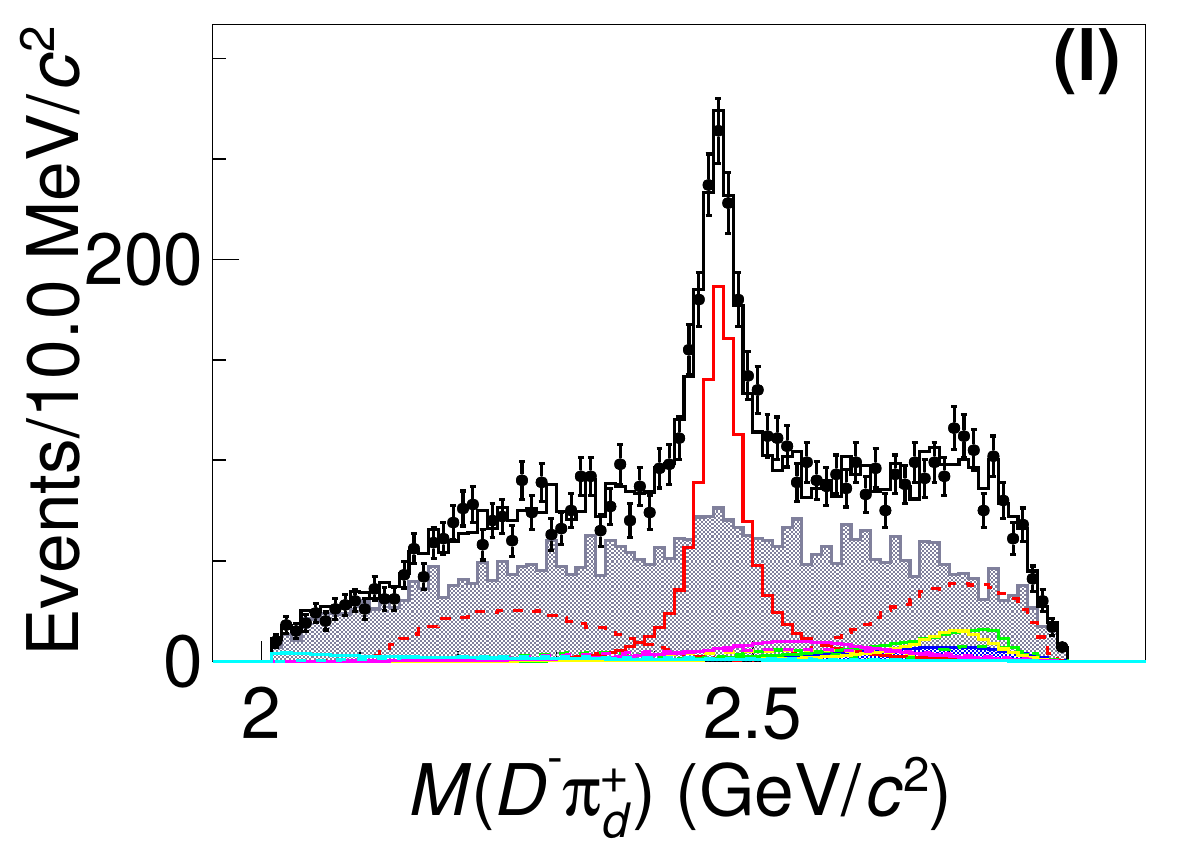}
    \caption{
	Distributions of $M(D^0D^-)$ (a), $M(D^0\pi_d^+)$ (b), and $M(D^-\pi_d^+)$ (c) in the DT sample.
	Distributions of $RM(\pi_d^+)$ (d), $M(D^0\pi_d^+)$ (e), and recoiling mass of $D^0$ [$RM(D^0)$] (f) in the ${\rm ST}^{0}_{K\pi}$ sample.
	Distributions of $RM(\pi_d^+)$ (g), $M(D^0\pi_d^+)$ (h), and $RM(D^0)$ (i) in the ${\rm ST}^{0}_{K3\pi}$ sample.
	Distributions of $RM(\pi_d^+)$ (j), $RM(D^-)$ (k), and $M(D^0\pi_d^+)$ (l) in the ${\rm ST}^{-}_{K2\pi}$ sample. 
	The black dots with error bars represent the $S$ sample, while the black histogram shows the total PWA fit projection.
	The blue histogram represents the $X(3790)^-$ component.
	The red, green, yellow, magenta, and cyan histograms correspond to the contributions from $D_2^*(2460)$, $D_3^*(2750)$, $D_1^*(2760)$, $D_1^*(2600)$, and $D^*$, respectively.
	Solid histograms indicate the charged states, whereas dashed ones correspond to the neutral states.
    The $\chi^{2}/n_{\rm bin}$ values from left to right and top to bottom are 1.38 (a), 0.99 (b), 1.09 (c), 0.99 (d), 1.00 (e), 1.47 (f), 0.97 (g), 0.90 (h), 1.38 (i), 1.13 (j), 1.41 (k), and 0.74 (l), respectively.
	}
    \label{pic:pwa_base}
\end{figure*}

Figure~\ref{pic:pwa_base} shows mass projections from PWA.

\begin{figure*}[!htbp]
    \centering
    \includegraphics[width=0.3\textwidth]{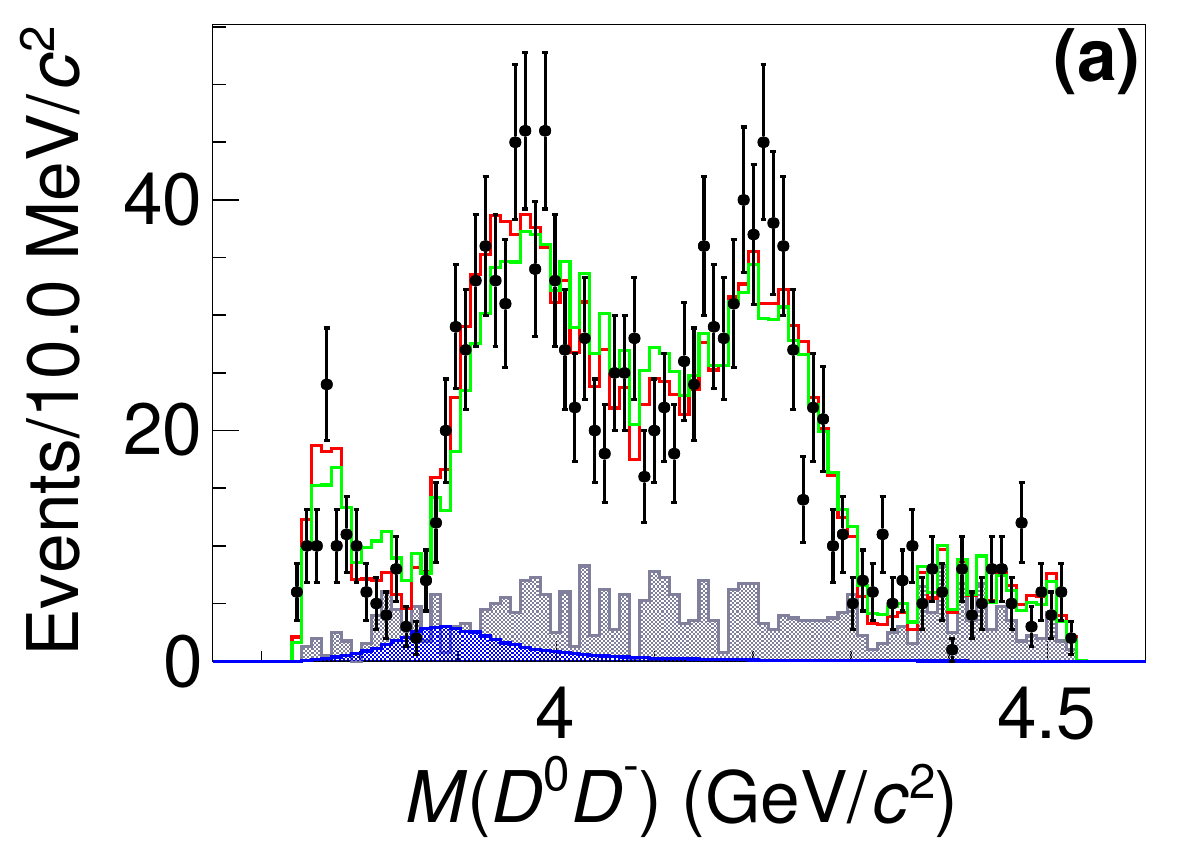}
    \includegraphics[width=0.3\textwidth]{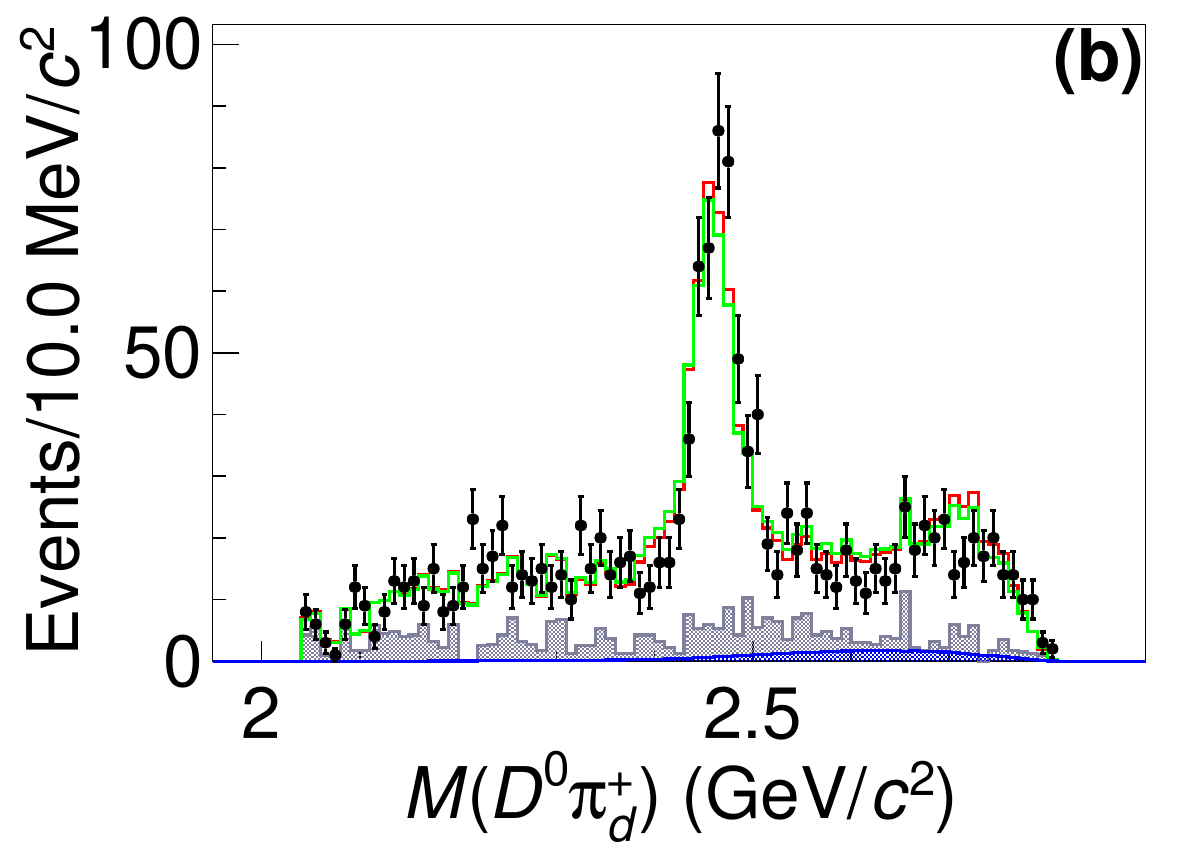}
    \includegraphics[width=0.3\textwidth]{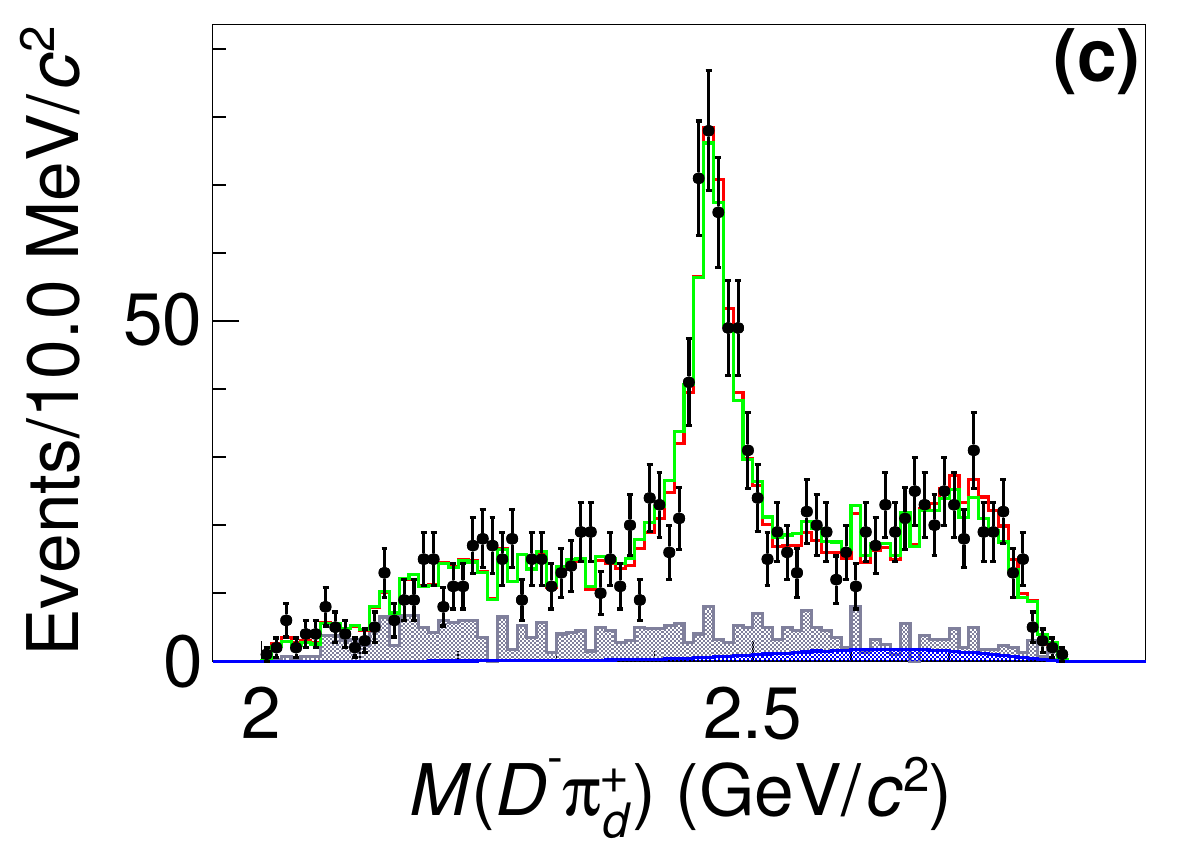}
    \includegraphics[width=0.3\textwidth]{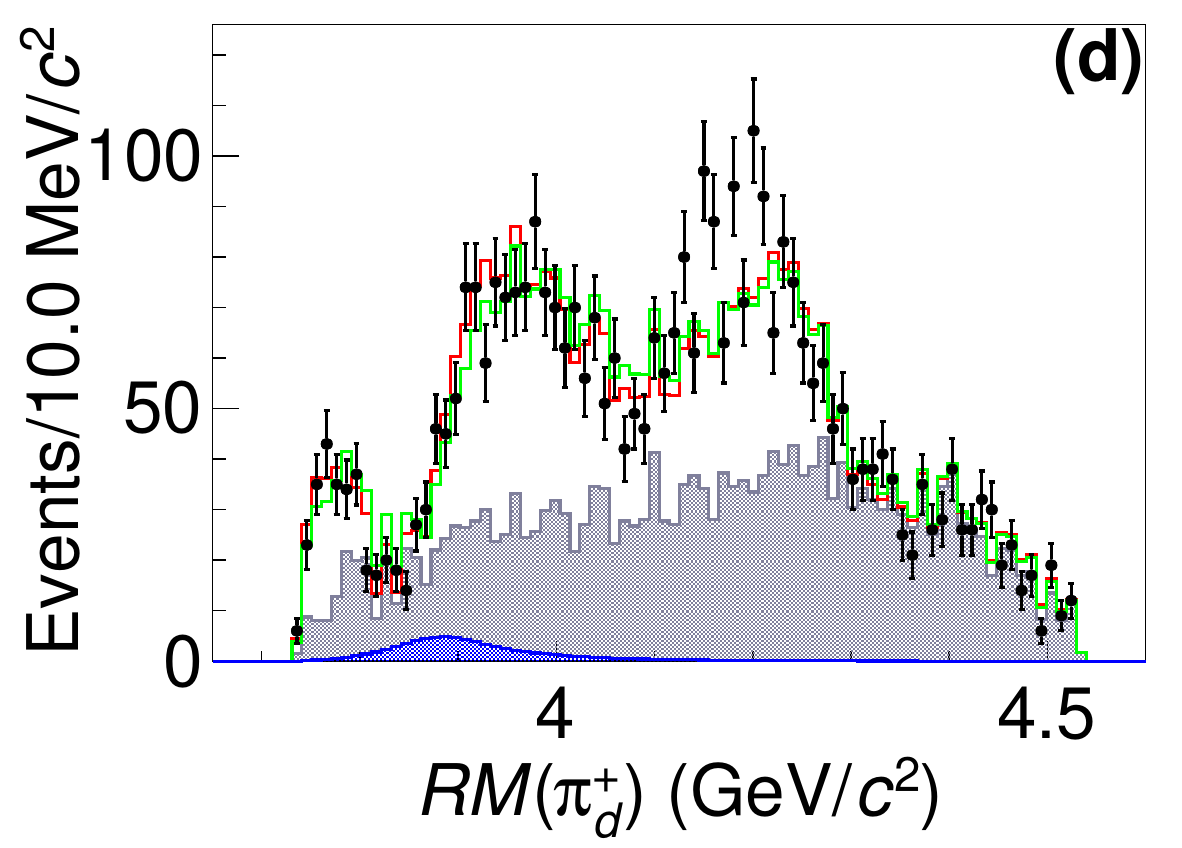}
    \includegraphics[width=0.3\textwidth]{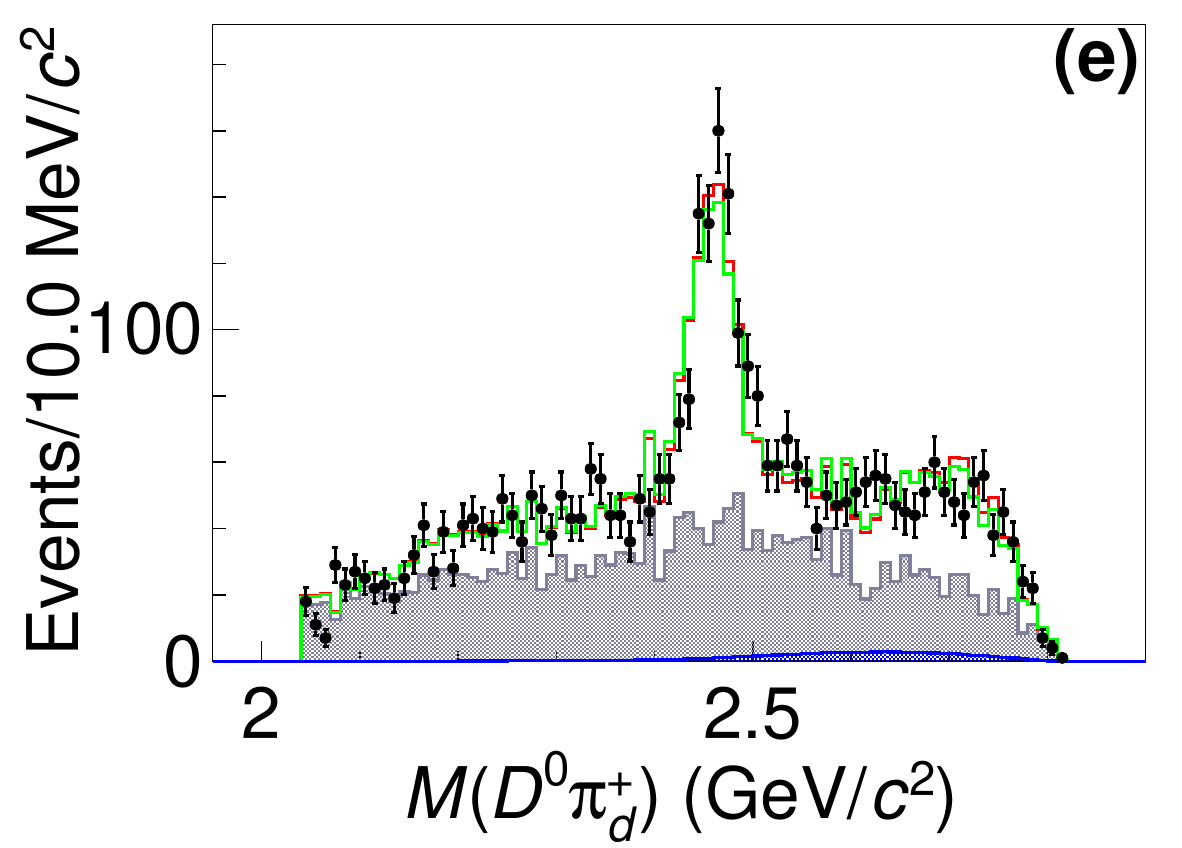}
    \includegraphics[width=0.3\textwidth]{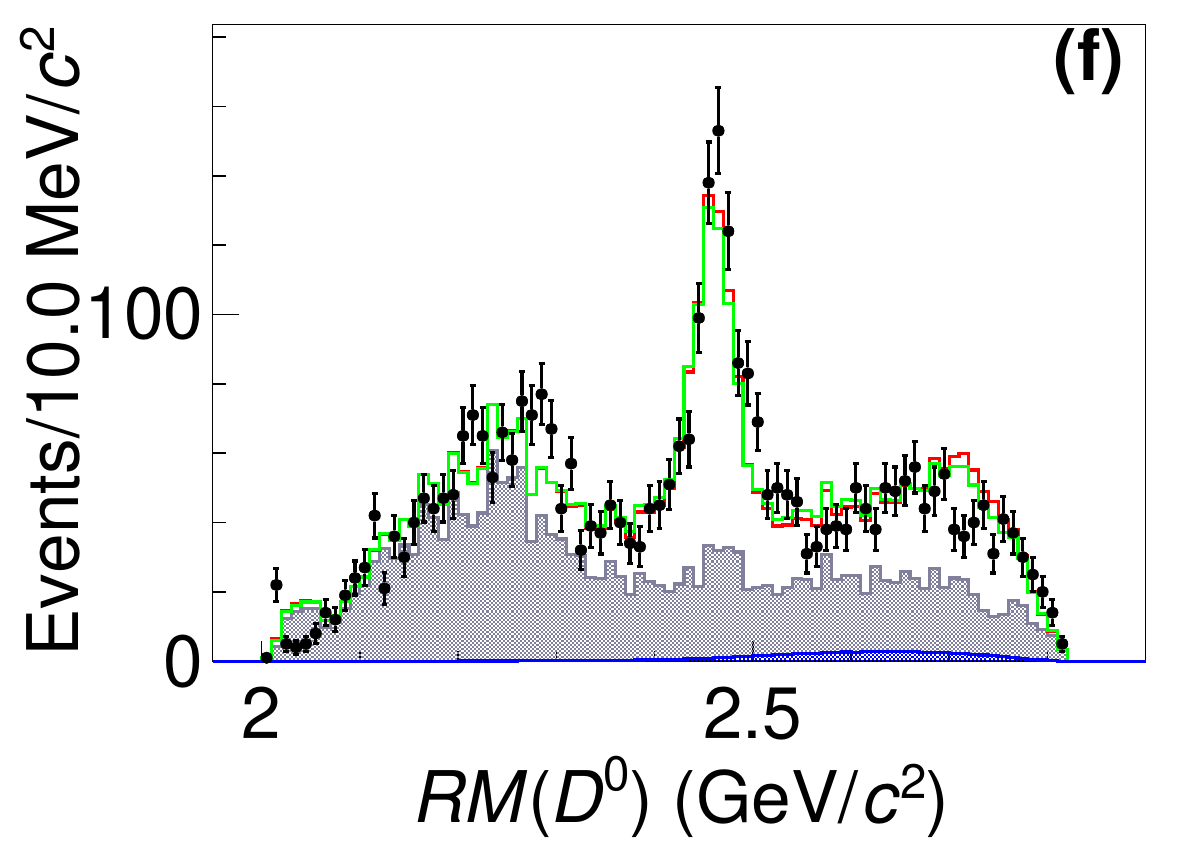}
    \includegraphics[width=0.3\textwidth]{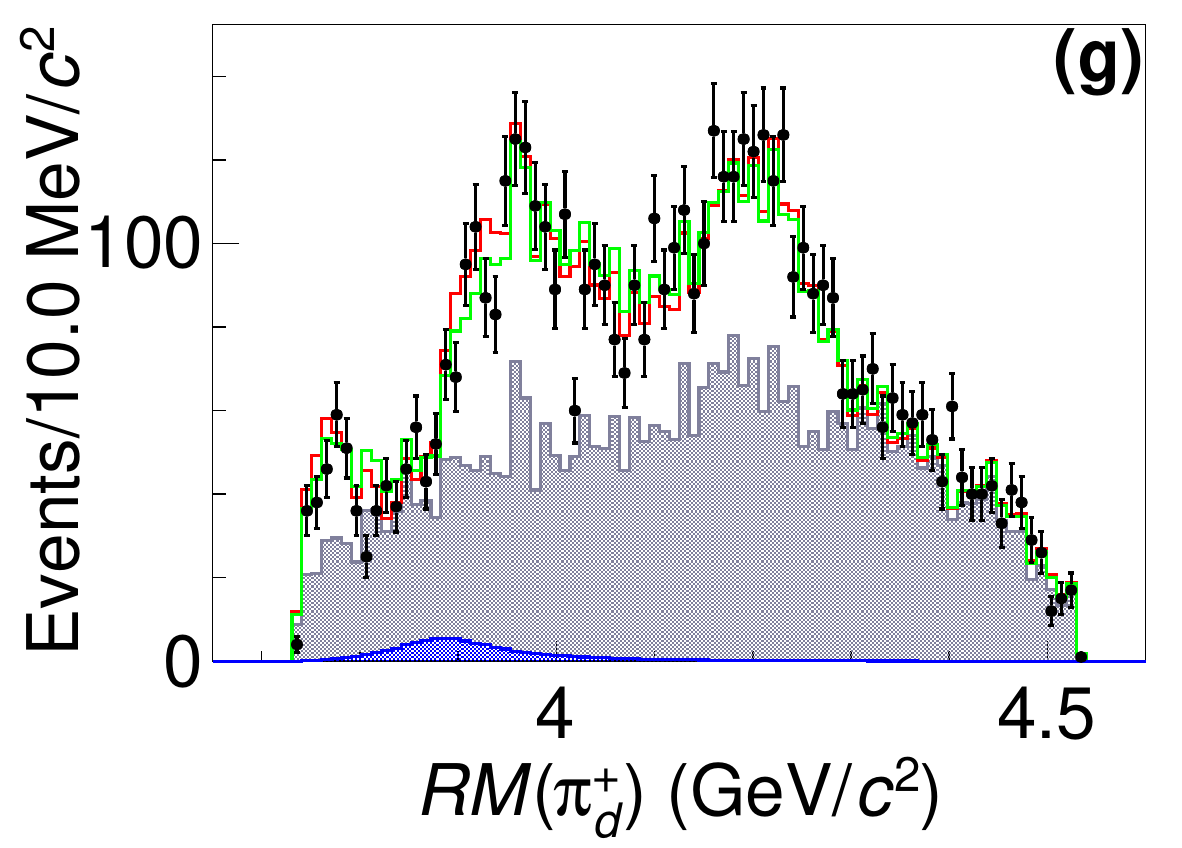}
	\includegraphics[width=0.3\textwidth]{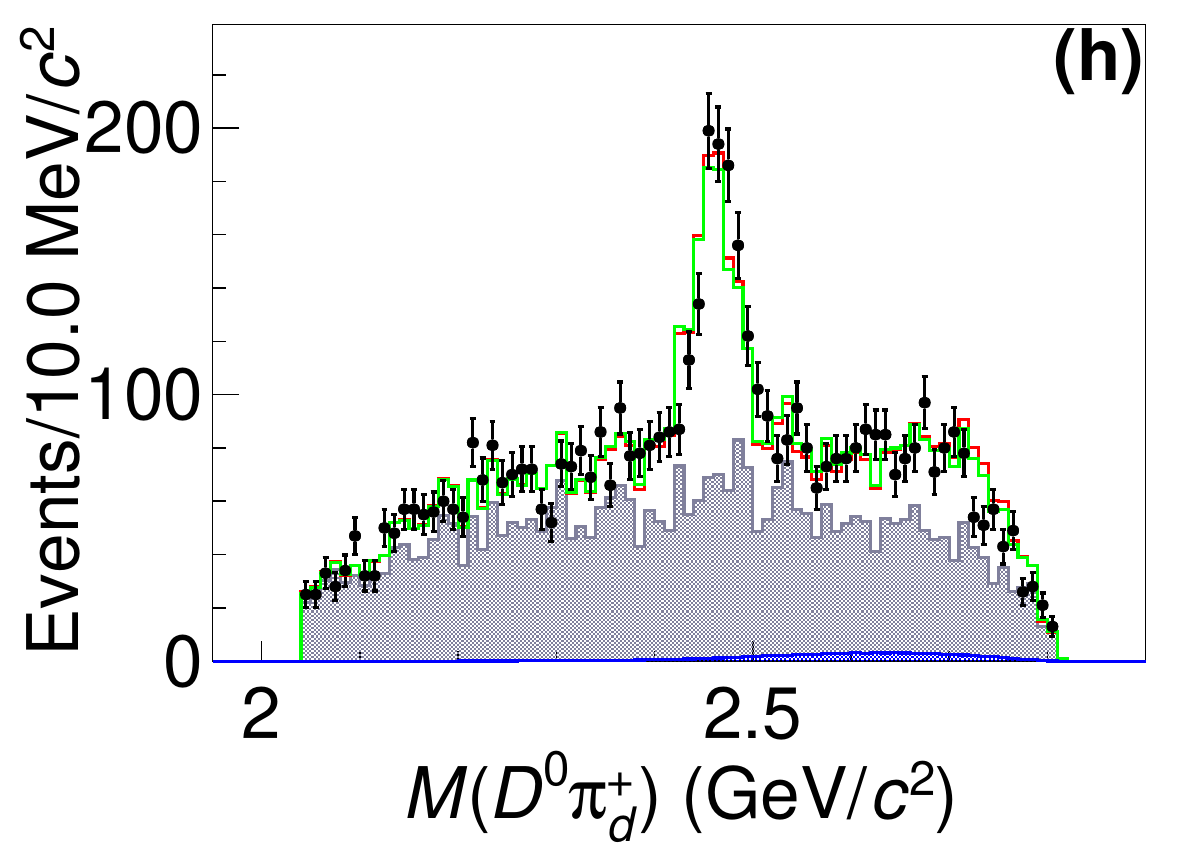}
	\includegraphics[width=0.3\textwidth]{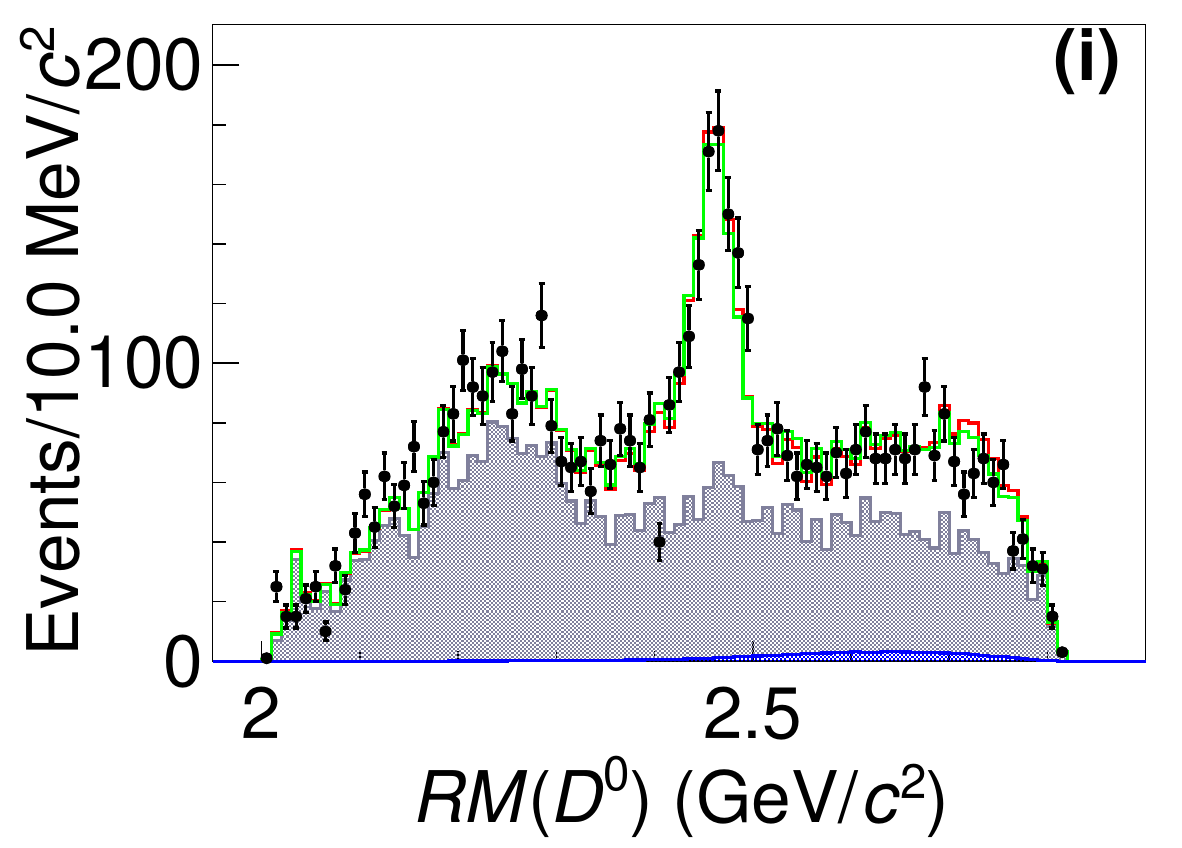}
    \includegraphics[width=0.3\textwidth]{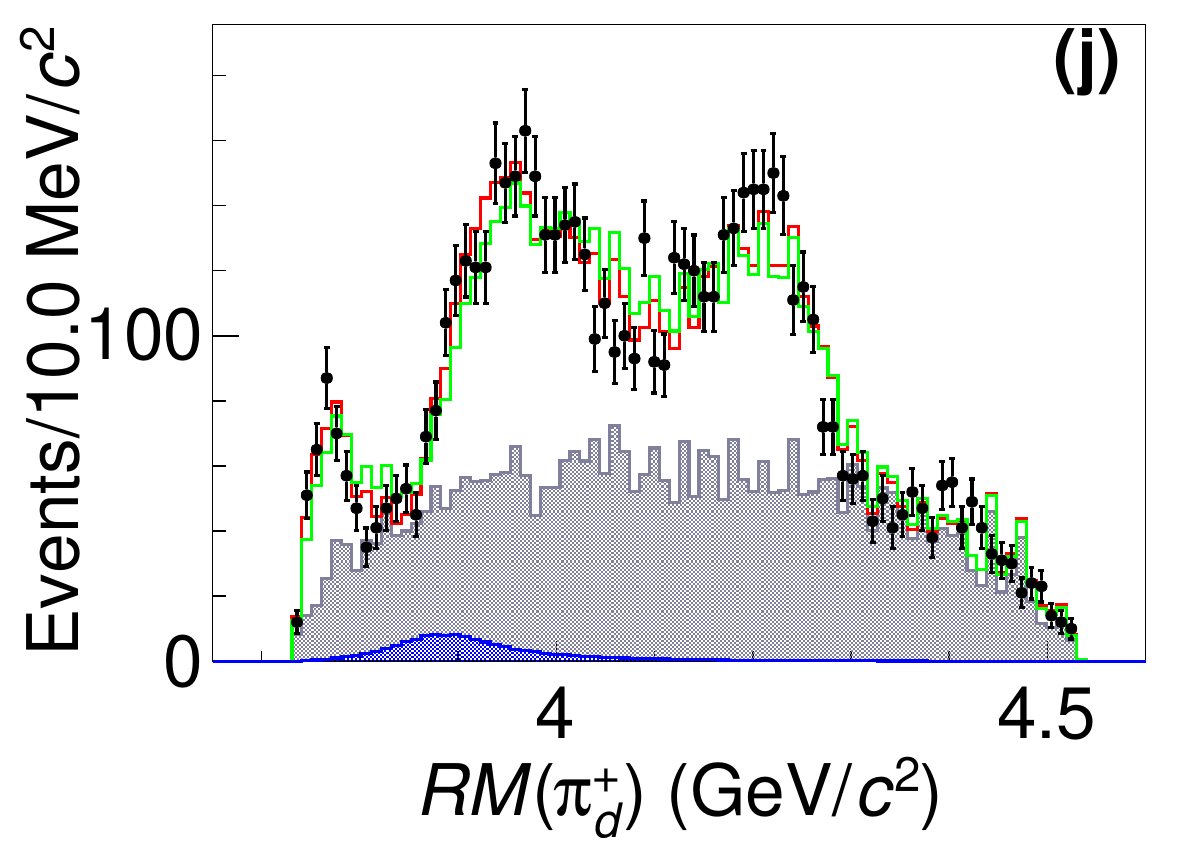}
	\includegraphics[width=0.3\textwidth]{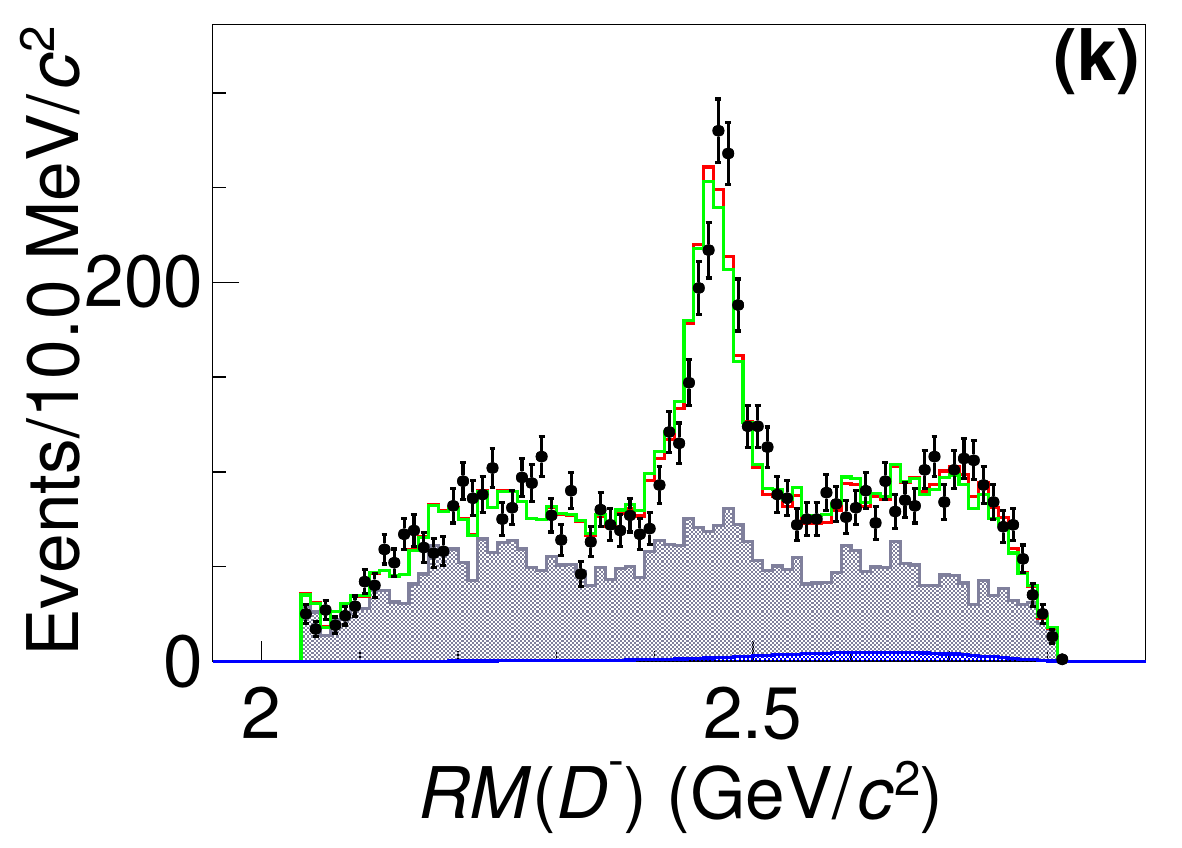}
	\includegraphics[width=0.3\textwidth]{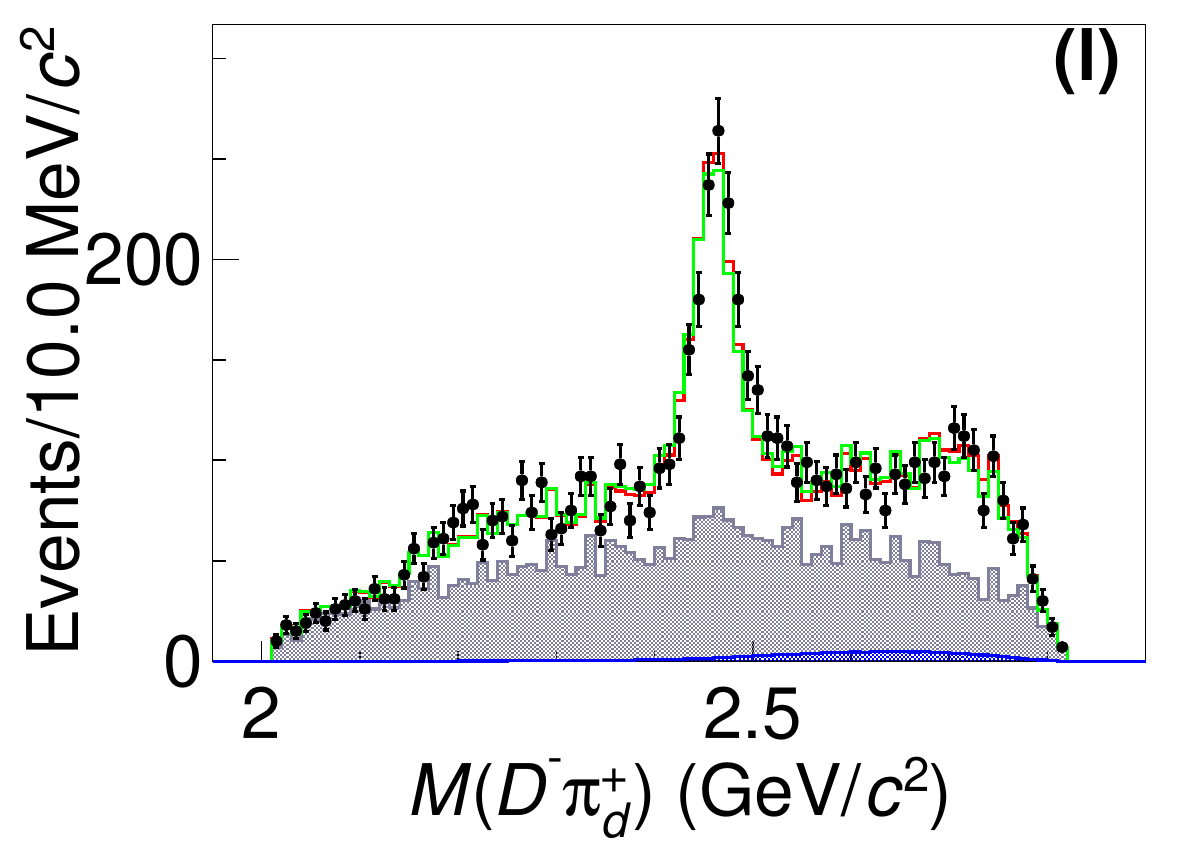}
    \caption{
	Distributions of $M(D^0D^-)$ (a), $M(D^0\pi_d^+)$ (b), and $M(D^-\pi_d^+)$ (c) in the DT sample.
	Distributions of $RM(\pi_d^+)$ (d), $M(D^0\pi_d^+)$ (e), and $RM(D^0)$ (f) in the ${\rm ST}^{0}_{K\pi}$ sample.
	Distributions of $RM(\pi_d^+)$ (g), $M(D^0\pi_d^+)$ (h), and $RM(D^0)$ (i) in the ${\rm ST}^{0}_{K3\pi}$ sample.
	Distributions of $RM(\pi_d^+)$ (j), $RM(D^-)$ (k), and $M(D^0\pi_d^+)$ (l) in the ${\rm ST}^{-}_{K2\pi}$ sample. 
	The black dots with error bars represent the $S$ sample.
	The red (green) histogram correspond to PWA fit projections with (without) the $X(3790)^-$ component.
    The grey and blue histograms are the $B$ sample and the $X(3790)^-$ component, respectively.
    With the inclusion of the $X(3790)^-$ component, the $\chi^{2}/n_{\rm bin}$ values from left to right and top to bottom are 1.31 (a), 1.31 (b), 0.94 (c), 1.05 (d), 1.10 (e), 1.67 (f), 1.00 (g), 1.07 (h), 1.43 (i), 1.27 (j), 1.73 (k), and 0.93 (l), respectively.
	}
    \label{pic:pwa_base_pdg_all}
\end{figure*}

Figure~\ref{pic:pwa_base_pdg_all} shows mass projections from PWA with $D_{2}^{*}(2460)$, $D_{1}^{*}(2600)$, $D_{1}^{*}(2760)$, and $D_{3}^{*}(2750)$ fixed to their nominal values in PDG~\cite{pdg}.

\section{Systematic Uncertainties}

(\romannumeral4) Background modeling: (1) To estimate the systematic uncertainty from the normalization factors in $B$ sample, we use a second order Chebyshev polynomial function and a Gaussian function to fit the distributions. 
In DT sample, the normalization factors obtained for $M(D^0)$ and $M(D^-)$ are 0.51 and 0.51, respectively;
in $\rm ST^0_{K\pi}$ sample, they are 0.57 and 0.51 for $M(K^-\pi^+)$ and $RM(D^0\pi_d^+)$, respectively;
in $\rm ST^0_{K3\pi}$ sample, they are 0.57 and 0.50 for $M(K^-\pi^+\pi^+\pi^-)$ and $RM(D^0\pi_d^+)$, respectively; and 
in $\rm ST^-_{K2\pi}$ sample, they are 0.53 and 0.50 for $M(K^+\pi^-\pi^-)$ and $RM(D^-\pi_d^+)$, respectively.
(2) The systematic uncertainty from background shape is estimated by shifting the sideband regions horizontally or vertically by 2~MeV/$c^2$ away from the signal region. 

(\romannumeral7) Detector resolution: (1) The systematic uncertainty from mass shift is estimated by the fits of $\ee\to \bar{D}_{2}^{*}(2460)^{0}D^{0}, \bar{D}_{2}^{*}(2460)^{0}\to D^{-}\pi^{+}$ MC sample (select the events as the nominal selection criteria). 
The signal shape is described by the line-shape when generating MC events convolved with a Gaussian function, the maximum central value (0.9~MeV/$c^{2}$) and standard deviation (0.3~MeV) of the Gaussian function in ${\rm DT}$, ${\rm ST}^{0}_{K\pi}$, ${\rm ST}^{0}_{K3\pi}$, and ${\rm ST}^{-}_{K2\pi}$ samples are taken as the systematic uncertainties for the masses and widths, respectively.
(2) The systematic uncertainty from the difference between data and MC samples in detector resolution are estimated by the fits of $M_1$ and $M_2$ with MC shape convolved with a Gaussian function as signal shape and a second order Chebyshev polynomial function as background shape, the maximum central value (0.1~MeV/$c^{2}$) and standard deviation (0.5~MeV) of the Gaussian function in ${\rm DT}$, ${\rm ST}^{0}_{K\pi}$, ${\rm ST}^{0}_{K3\pi}$, and ${\rm ST}^{-}_{K2\pi}$ samples are taken as the systematic uncertainties for the masses and widths, respectively. 
The systematic uncertainty from detector resolution is taken as 0.9~MeV$/c^{2}$ and 0.6~MeV for all the masses and widths, respectively. 
Especially, the systematic uncertainty for $g$ is estimated by $\frac{{\rm sys.\ unt.}_{\Gamma_{D_{2}^{*}(2460)^+}}}{\Gamma_{D_{2}^{*}(2460)^+}}\times g=\frac{0.6}{40.2}\times1.0 {\rm\ GeV} \approx0.0$ GeV, where ${\rm sys.\ unt.}_{\Gamma_{D_{2}^{*}(2460)^+}}$ is the systematic uncertainty on $\Gamma_{D_{2}^{*}(2460)^+}$ from detector resolution.

The total systematic uncertainties, as summarized in Table~\ref{tab:sys_tot}, are obtained by summing all individual sources in quadrature under the assumption that they are uncorrelated.

\begin{table*}[htp]
    \centering
    \caption{Systematic uncertainties on parameters of $D^{*}_{2}(2460)^{+}$, $D^{*}_{2}(2460)^{0}$, $X(3790)^{-}$, $D^{*}_{1}(2760)$, $D^{*}_{1}(2600)$, and $X(3790)^-$. The units for mass, width, and $g$ are MeV/$c^{2}$, MeV, and GeV, respectively. The symbol ``-'' denotes that the value is zero after rounding to the number of efficient significant digits.}
    \begin{tabular}{cccccccccccc}
    \hline\hline\noalign{\smallskip}
     & $m_{D^{*}_{2}(2460)^{+}}$  & $\Gamma_{D^{*}_{2}(2460)^{+}}$   & $m_{D^{*}_{2}(2460)^{0}}$  & $\Gamma_{D^{*}_{2}(2460)^{0}}$  & $m_{X(3790)^{-}}$ & $g_{0}$ & $m_{D^{*}_{1}(2760)}$ & $\Gamma_{D^{*}_{1}(2760)}$ & $m_{D^{*}_{1}(2600)}$ & $\Gamma_{D^{*}_{1}(2600)}$ \\
    \noalign{\smallskip}\hline\noalign{\smallskip}
    \romannumeral1 & 1.0 & 0.2 & 0.7 & 0.7 & 8.1 & 0.0 & 2.2 & 1 & 8 & 10 &\\
    \romannumeral2 & 0.1 & - & 0.1 & - & 0.1 & - & 0.1 & - & - & 1 &\\
    \romannumeral3 & 1.6 & 0.1 & 1.0 & 1.4 & 2.3 & 0.2 & 4.9 & 18 & 20 & 36 &\\
    \romannumeral4 & 0.7 & 0.3 & 0.4 & 2.2 & 1.7 & 0.2 & 2.9 & 2 & 1 & 5 &\\
    \romannumeral5 & - & - & - & - & - & - & - & - & - & - &\\
    \romannumeral6 & 0.2 & 0.1 & 0.1 & 0.1 & 0.4 & - & - & - & 1 & 1 &\\
    \romannumeral7 & 1.3 & 0.6 & 1.3 & 0.6 & 1.3 & - & 1.3 & 1 & 1 & 1 &\\
    Total & 2.4 & 0.7 & 1.8 & 2.8 & 8.7 & 0.2 & 6.3 & 18 & 22 & 38 &\\
    \noalign{\smallskip}\hline\hline
    \end{tabular}
    \label{tab:sys_tot}
\end{table*}